\documentclass[12pt]{iopart}

\usepackage{bm}
\usepackage{graphicx}
\usepackage{color}
\usepackage{subfigure}
\usepackage{hyperref}
\usepackage{latexsym}
\usepackage{amsthm}
\usepackage{amssymb}
\usepackage{braket}
\usepackage{cite}
\usepackage[rgb]{xcolor}
\usepackage{hyperref}

\usepackage{bbold}
\usepackage{bbm}
\usepackage{verbatim}
\usepackage{multirow}
\usepackage[english]{babel}
\usepackage{comment}
\usepackage{bm}
\DeclareGraphicsExtensions{.jpg,.pdf, .mps, .png, .eps, .ps, .EPS,.gif}
\usepackage{url}

\DeclareMathAlphabet{\mathbcal}{OMS}{cmsy}{b}{n}
\begin{document}

\def\be{\begin{equation}}
\def\ee{\end{equation}}

\def\bc{\begin{center}}
\def\ec{\end{center}}
\def\bea{\begin{eqnarray}}
\def\eea{\end{eqnarray}}
\newcommand{\avg}[1]{\langle{#1}\rangle}
\newcommand{\Avg}[1]{\left\langle{#1}\right\rangle}

\def\ie{\textit{i.e.}}
\def\etal{\textit{et al.}}
\def\m{\vec{m}}
\def\G{\mathbcal{G}}

\newcommand{\davide}[1]{{\bf\color{blue}#1}}
\newcommand{\gin}[1]{{\bf\color{green}#1}}
\title[The topological Dirac equation of  networks and simplicial complexes]{The topological Dirac equation of  networks and simplicial complexes}

\author{Ginestra Bianconi}

\address{School of Mathematical Sciences, Queen Mary University of London, London, E1 4NS, United Kingdom\\
Alan Turing Institute, The British Library, London, United Kingdom}
\ead{ginestra.bianconi@gmail.com}
\vspace{10pt}
\begin{indented}
\item[]
\end{indented}

\begin{abstract}
We define the topological Dirac equation describing the evolution of a topological wave function  on  networks or on simplicial complexes. On networks, the topological wave function describes the dynamics of topological signals or cochains, i.e. dynamical signals   defined both on  nodes and on  links. On simplicial complexes the wave function is also  defined  on   higher-dimensional simplices. Therefore the topological wave function satisfies a relaxed condition of locality as it acquires the same value along simplices of dimension larger than zero. 
The topological Dirac equation   defines eigenstates whose dispersion relation is determined by the spectral properties of the Dirac (or chiral) operator defined on networks and generalized network structures including   simplicial complexes and multiplex networks.  On simplicial complexes the Dirac equation leads to  multiple energy bands. On multiplex networks the topological Dirac equation can be generalized to distinguish between different mutlilinks leading to a natural definition of rotations of the topological spinor. The topological Dirac equation is here initially  formulated on a spatial network or simplicial complex for describing the evolution of the topological wave function in continuous  time. This framework is  also extended to  treat the topological Dirac equation on $1+d$ spaces describing a discrete space-time with one temporal dimension and $d$ spatial dimensions with $d\in \{1,2,3\}$.  This  work includes also the  discussion  of  numerical results obtained by implementing the topological Dirac equation on  simplicial complex  models   and on real simple and multiplex network data.
\end{abstract}

%
%
%
%
%

\section*{Introduction}

Topology \cite{hatcher,jost2015mathematical,ghrist2008barcodes,giusti2016two,otter2017roadmap}
is emerging as a very powerful tool to reveal important structural properties of interacting networks and simplicial complexes. Indeed persistence homology \cite{patania2017topological,petri2014homological} can reveal large scale-properties of data that cannot be captured by more traditional network measures that characterize  mostly the combinatorial properties of networks. 
New results  \cite{millan2020explosive,torres2020simplicial,reitz2020higher,bick2021higher,mulas2020coupled,jost2019hypergraph,barbarossa2020topological,schaub2020random,ebli2020simplicial,bodnar2021weisfeiler,taylor2015topological,battiston2020networks}      are showing that topology can also greatly enrich our understanding  of  dynamics. In fact networks   can sustain topological signals that are dynamical variables not only defined on the nodes but also on the links. For instance topological signals defined on links, or 1-cochains, can model fluxes associated to links. On simplicial complexes, which are discrete interacting structures not only formed by nodes and links but also  by  higher-dimensional simplices,  topological signals can be  defined on simplices of every dimension including nodes, links, triangles, tetrahedra and so on.
By combining algebraic topology and dynamics, important progress has been made in capturing the higher-order synchronization dynamics  and the higher-order diffusion dynamics of topological signals \cite{millan2020explosive,torres2020simplicial,reitz2020higher}. Moreover progress as been made in  formulating signal processing techniques to analyse data in the form of topological signals \cite{barbarossa2020topological,schaub2020random,ebli2020simplicial,bodnar2021weisfeiler}

Topology is receiving  a surge of interest also in theoretical physics \cite{yang2019topology} and is increasingly used in high energy physics, quantum gravity and in condensed matter as well. In particular topology reveals important properties of new states of matter such as topological insulators \cite{shen2012topological} and higher-order superconductors \cite{mazziotti2021multigap}. Interestingly quantum algorithms \cite{lloyd2016quantum} have also been proposed to speed up Topological Data Analysis.

An important question is whether topology can also be explored to investigate the properties of quantum complex networks \cite{biamonte2019complex}.

Despite  quantum complex network is a term used to indicate very different approaches, all approaches that go under this name combine network and graph theory with quantum mechanics in order to capture non-trivial effects induced by  the architecture of complex networks on quantum dynamics. We can broadly classify the approaches proposed so far in two major classes. The first class of models of quantum complex networks interpret network as ``physical systems". This approach includes the study of quantum critical dynamics defined on complex networks instead of the traditionally investigated lattices\cite{bianconi2012superconductor,halu2012phase}, or quantum dynamics on non-commutative geometry of graphs \cite{majid2013noncommutative}, and approaches that associate quantum-oscillators to the nodes of a network and couple them to a single probe  whose frequency can be modulated freely\cite{nokkala2016complex,nokkala2018reconfigurable,nokkala2020probing}.  
The second class of approaches uses networks as ``abstract computational objects" that  can represent  quantum states or  the entanglement present in many-body systems \cite{hein2004multiparty,rossi2013quantum,anand2011shannon,bianconi2017emergent,de2016spectral,garnerone2012bipartite,valdez2017quantifying,sundar2018complex,bianconi2016network}. This has lead to the definition of Von Neumann entropy of networks \cite{anand2011shannon}, and its generalizations based on the spectral properties of networks \cite{de2016spectral}. This class of models includes also models of networks \cite{bianconi2001bose} and simplicial complexes \cite{bianconi2016network,bianconi2017emergent}  whose statistical properties are described by quantum statistics. Finally  the formulation of complex networks built from weighted matrices of mutual information characterizing  interacting many-body systems \cite{valdez2017quantifying,sundar2018complex} follows also in this class of models.

In this work our aim is to use algebraic topology to investigate the interplay between topology and the evolution of wave functions. We  formulate  the topological Dirac equation  describing the dynamics of a topological wave function.
 On networks, the topological Dirac equation   acts on a discretized wave function which is defined on each node and each link, i.e. is the combination of different topological signals (a 0-cochain and a 1-cochain) associated to the network. 
This implies that the discretized wave function includes dynamical variables that are defined also on links, i.e. we are ready to sacrifice slightly the notion of locality of the wave function.
We have therefore that the whole wave function of the network is defined by a vector here called  {\em topological spinor} which is formed by two blocks: one characterizing the value of the wave function on the nodes one characterizing the value of the wave function on the links of the network.
The topological Dirac equation is then simply written by appropriately using the discrete Dirac operator (also known as chiral operator) associated to the network. We note that while in discrete topology and discrete geometry the name of Dirac operator is used to indicate different operators such as the one used in Ref. \cite{crane2011spin}, here we leverage on the definition adopted in Ref.\cite{lloyd2016quantum} to propose quantum algorithms for speeding up Topological Data Analysis.

The topological Dirac equation determines a class of  quantum complex networks that  uses networks both as a physical system indicating the substrate of the quantum dynamics (as the wave function is defined on the network) and as computational representation of the wave function (as the wave function is a combination of 0-cochains and 1-cochains defined on networks). In some sense the topological Dirac equation goes in the opposite direction of quantum graphs\cite{berkolaiko2013introduction,kurasov2005inverse,kennedy2016spectral
}: while quantum graphs describe piecewise continuous wave functions obeying the linear Schr\"oedinger equation on each link, the topological Dirac equation describes a completely discretized wave-function taking a given (constant) value on each   node and on each link.

Our aim is to explore the relation between the spectral properties of the network and the eigenstates of the topological Dirac equation.

The spectral properties of networks \cite{chung1997spectral,burioni2005random,rammal1983random} are known to be fundamental to investigate diffusion on networks and simplicial complexes \cite{torres2020simplicial,reitz2020higher,schaub2020random,jost2019hypergraph} and they  provide a natural link between topology and dynamics. Characterizing the spectral properties of graphs and networks constitute and important branch of graph theory \cite{chung1997spectral}, network theory \cite{burioni2005random,rammal1983random} and machine learning\cite{hamilton2020graph}. Interestingly the spectral properties of networks have recently attracted also increasing attention in the quantum gravity community and in the field of quantum networks \cite{calcagni2013probing,benedetti2009fractal,calcagni2014spectral,nokkala2020probing}.

Here we reveal the relation between the topological Dirac equation and the spectral properties of simple networks,  simplicial complexes  and multiplex networks.

While networks are only objects formed by nodes and links generalized network structures including  simplicial complexes and multilayer networks are able to encode more complex interacting patterns.  Simplicial complexes \cite{battiston2020networks,bick2021higher} capture the many body interactions of a complex systems including not only pairwise interactions (like networks) but also higher-order interactions such as  three-body interactions (triangles) and four-body interactions (tetrehedra) and so on. Multilayer networks \cite{bianconi2018multilayer} are able to capture network of networks describing complex systems in which a given set of nodes is related by different types of interactions. As the vast majority of complex systems from the brain to technological and transportation network is formed by elements (nodes) connected via interactions that have different nature and connotation, multilayer networks are ubiquitous when describing complex systems.

The spectral properties of simplicial complexes \cite{torres2020simplicial,reitz2020higher,schaub2020random} and multilayer networks \cite{radicchi2013abrupt} are attracting great interest in recent years.
Here we show how their spectral properties affects the topological Dirac equation. On networks the topological Dirac equation describes the dynamics of  wave functions with relativistic dispersion relation in which the role of the momentum is played by the eigenvalue of the graph Laplacian. On  simplicial complexes the topological Dirac equation has a dispersion relation that includes more than a single energy band with each energy band determined by the spectral properties of the $n$th order up-Laplacian. The eigenstates corresponding to the eigenvalues in the $n$-th energy band are also localized exclusively on $n$ and $n+1$ simplices.

The topological Dirac equation defined on networks can be  compared with the Dirac equation in one dimension, and in a lattice  does not distinguishes links according to their directionality. In order to distinguish between $x$-links, $y$-links and $z$-links in lattices of dimension $d=2$ and $d=3$ we introduce here the directional topological Dirac equation. While several approaches to quantum gravity ranging from spin networks to spin foams \cite{penrose1971angular,rovelli1995spin,rovelli2014covariant,baez1998spin,oriti2001spacetime} associate spins to  links of the network here the approach is developed in the simple case in which the directions of the links are orthogonal leading to a very simplified treatment of the rotations induced by the different directions of the links.

Inspired by the Dirac equation in two and three dimension we can formulate a directional topological Dirac equation for duplex networks, (i.e. multiplex networks formed by two layers). 
This directional topological Dirac equation is able to discriminate between the different types of interactions connecting any two pair of nodes encoded by the so called multilinks \cite{bianconi2013statistical}. This implies that the directional topological Dirac equation will treat differently multilinks of types $(1,0),(0,1)$ and $(1,1)$ characterizing the interaction of pair of nodes connected only in layer 1, only in layer 2 or in both layers respectively.  This approach  leads to a natural definition of a  topological spinor formed by two $0$-cochains and two $1$-cochains and naturally associates a different rotation of the topological spinor to   different types of multilinks. 

All the above results are obtained for the topological Dirac equation or its directional generalizations which are acting  on a topoloigcal spinor defined on a network  describing the spatial topology, while time is treated as a continuous variable.
To address the interesting problem whether the topological Dirac equation can be defined on a discretized space-time. Here we treat the illustrative case of a $1+d$ dimensional lattice having $d\in \{1,2,3\}$ spatial dimension and one temporal dimension. Interestingly we can treat the space as a simple Euclidean space in $\mathbb{R}^{d+1}$ dimension while the directional Dirac operator can be defined differently for different directions $w$. In particular choosing the spatial directional Dirac operators as Hermitian and the temporal directional Dirac operator as an suitably defined anti-Hermitian matrix allows us to obtain eigenstates whose component defined on the links is an eigenvector of the discrete d'Alembert operator. 

The paper is defined as follows: In Sec.~1 we introduce the topological Dirac equation on general network structures; in Sec.~2 we show how the topological Dirac equation can be generalized for higher-order simplicial complexes, leading to a multi-band spectrum; in Sec.~3 we discuss the directional topological Dirac equation on lattices of dimension $d=2,3$; in Sec.~4 we show how the directional Dirac equation can be applied to  duplex networks in which we wish to distinguish between different types of multilinks; in Sec.~5 we  show how the topological Dirac equation can be applied to a $1+d$ dimensional space-time with $d\in \{1,2,3\}$. Finally in Sec.~6 we provide the concluding remarks.

\section{Topological Dirac equation of a network}
\subsection{Topological Dirac equation on a arbitrary network}
In this section we  introduce the   topological Dirac equation defined on networks. To this end  we consider a network $G=(V,E)$ formed by a set $V$ of $N$ nodes (or vertices) $\{1,2,\ldots, N\}$ and a set $E$ of $M$  links (or edges) $\{\ell_1,\ell_2,\ldots, \ell_M\}$ connecting the nodes by pairwise interactions.
Here we assume that the network $G$  has an arbitrary topology that might include chains and square lattices but can be used to treat also complex network topologies capturing the complexity of interacting systems ranging from the Internet to brain networks.

\begin{figure}[!htbp]
\begin{center}
\includegraphics[width=1.0\textwidth]{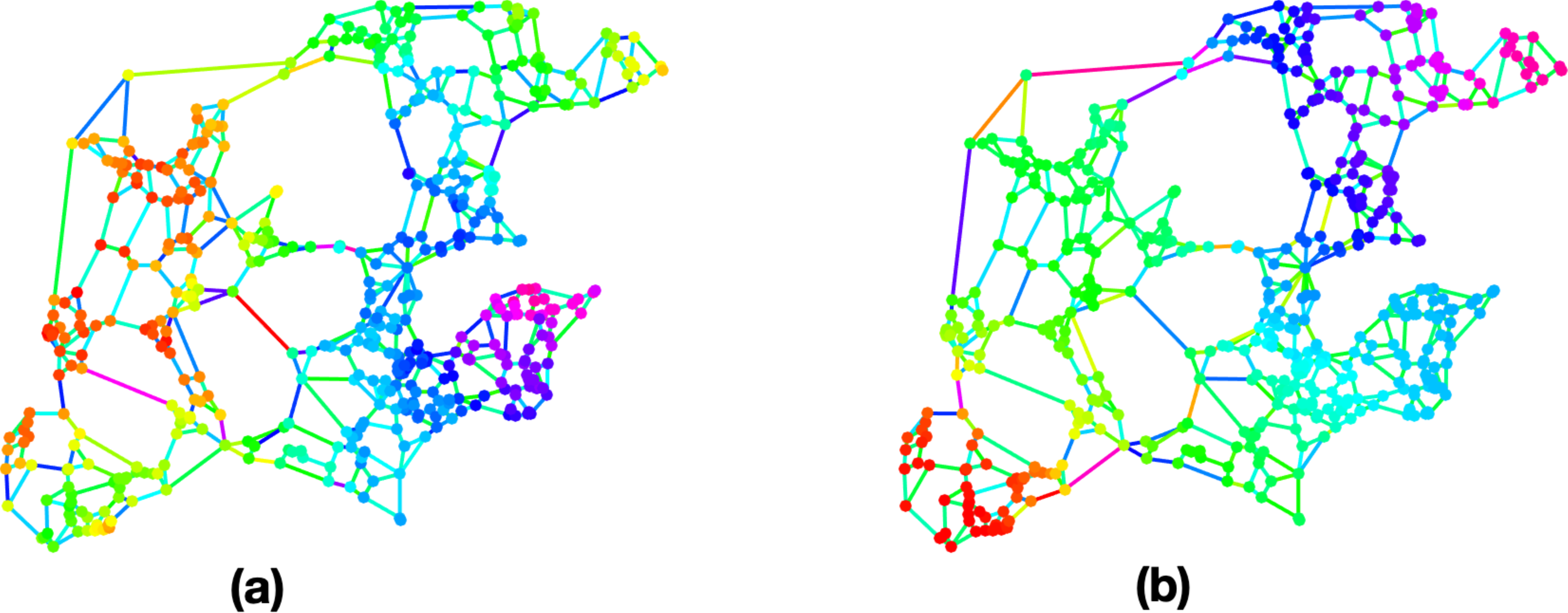}
\caption{{\bf Visualization of two eigenstates of the topological Dirac equation on a real network.} We represent two eigenstates of the topological Dirac equation on the fungal network ''PpMTokyoUN26h6" from Ref.\cite{Lee_2016_CP} (data available at \cite{fungi}). Here the topological Dirac equation is studied for $b=1$ and unitary mass $m_0=1$. The eigenstates represented in the two panels correspond to energy  $E=1.0033$ (panel a) and energy $E=-1.0038$ (panel b).}
\label{fig:1}
\end{center}
\end{figure}

\begin{figure}[!htbp]
\begin{center}
\includegraphics[width=1.0\textwidth]{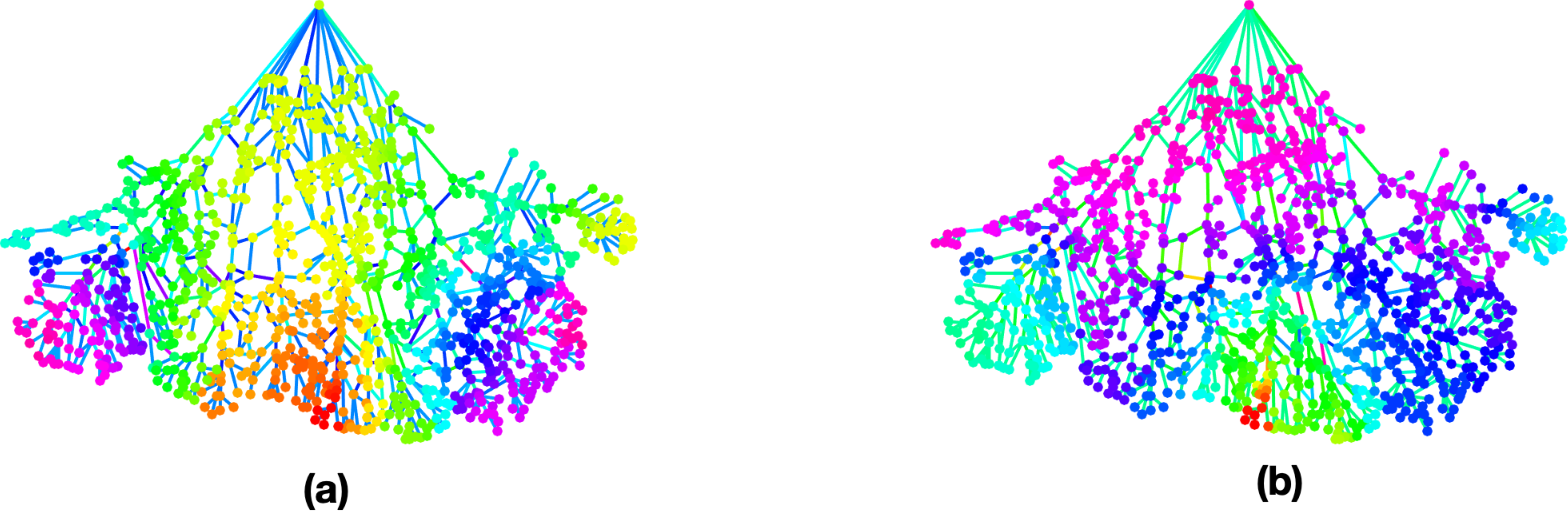}
\caption{{\bf Visualization of two eigenstates of the topological Dirac equation on a real network.} We represent two eigenstates of the topological Dirac equation on the fungal network ''PiMIUN46d6" from Ref.\cite{Lee_2016_CP} (data available at \cite{fungi}). Here the topological Dirac equation is studied for $b=1$ and unitary mass $m_0=1$. The eigenstates represented in the two panels correspond to energy  $E=1.0022$ (panel a) and energy $E=-1.0041$ (panel b).}
\label{fig:2}
\end{center}
\end{figure}

\begin{figure}
\begin{center}
\includegraphics[width=0.80\textwidth]{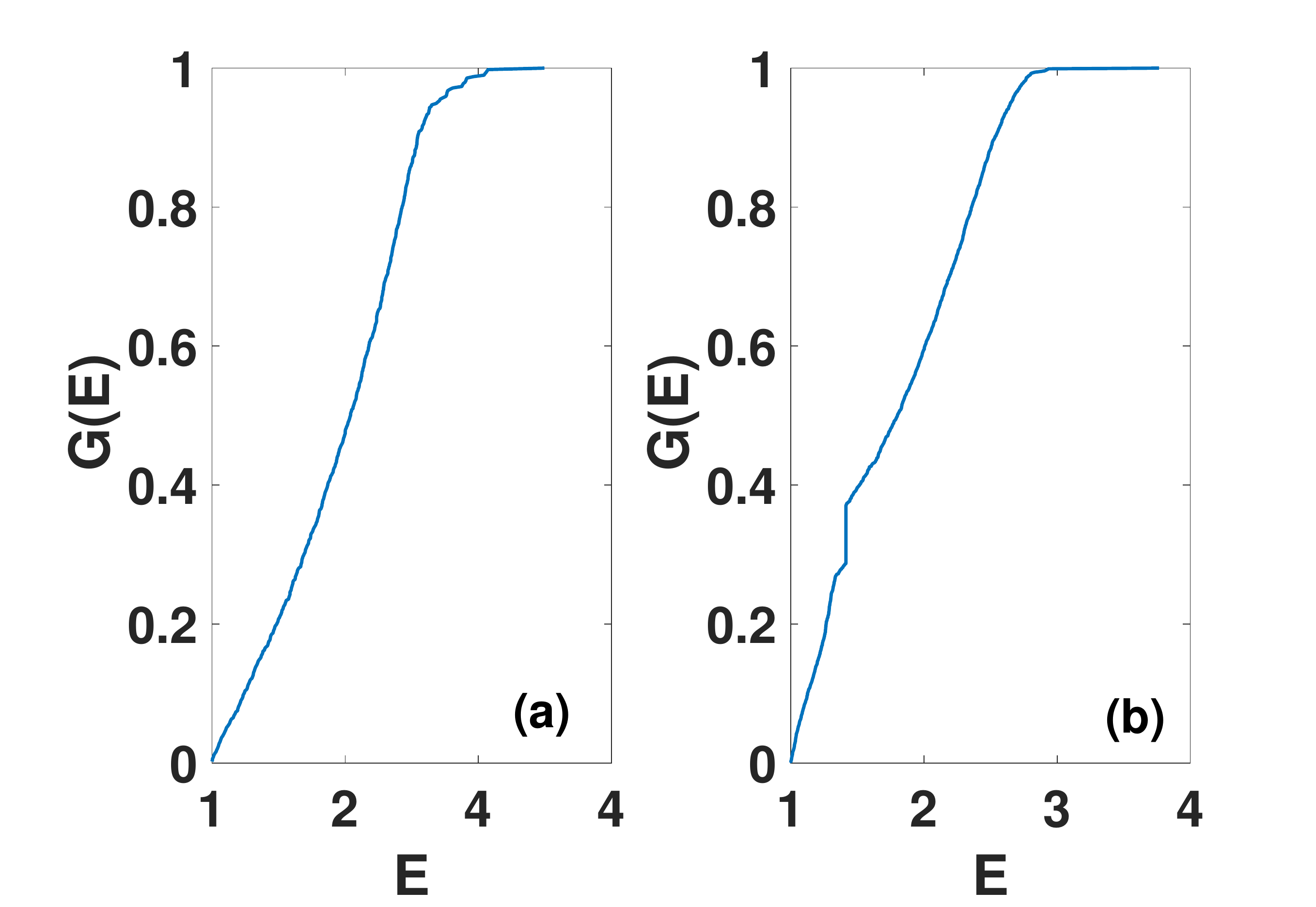}
\caption{ {\bf Cumulative density of  eigenstates of positive (negative) energy $E$ for real networks.} The cumulative density $G(E)$ of eigenstates  of positive energy $E$ with $E>m_0$, is shown for the fungal network displayed in Figure 1 (panel a) and the fungal network displayed in Figure 2 (panel b).
}
\label{fig:3}
\end{center}
\end{figure}

The topology of the network \cite{hatcher} can be captured by the incidence matrix ${\bf B}_{[1]}$ representing the boundary operator of the network mapping each link $\ell$ of the network to a linear combination of its two endnodes. In particular the incidence matrix ${\bf B}_{[1]}$ of a network (indicated here for ease of notation simply by ${\bf B}$)   is a rectangular matrix of size $N\times M$   having elements given by 
\bea
{\bf B}_{i\ell}=\left\{\begin{array}{ccc}1 & \mbox{if}\   \ell=[j,i],\\ -1 & \mbox{if} \ \ell=[i,j],\\  0 & \  \mbox{otherwise}.\end{array}\right.
\label{B_link}
\eea
The incidence matrix ${\bf B}$ is known to define the graph Laplacian ${\bf L}_{[0]}$ describing diffusion from node to node through links and the $1$-down-Laplacian ${\bf L}_{[1]}^{down}$ describing diffusion from link to link though nodes: 
\bea
{\bf L}_{[0]}={\bf B}{\bf B}^{\dag},\label{L0}\\
{\bf L}_{[1]}^{down}={\bf B}^{\dag}{\bf B}.
\eea
The Dirac operator (also called the  chiral operator) \cite{lloyd2016quantum} of a network is defined starting from the incidence matrix of the network and can be represented by  a square $(N+M)\times (N+M)$ matrix given by 
\bea
\mathbcal{D}=\left(\begin{array}{cc}0& b{\bf B}\\b^{\star}{\bf B}^{\dag}& 0\end{array}\right),
\eea
where  $b\in \mathbb{C}$ has  absolute value $|b|=1$ and where $b^{\star}$ is its complex conjugate. Typically we can choose $b$ equal to the real unit ($b=b^{\star}=1$) or $b$ equal to the imaginary unit ($b=-b^{\star}=\mathrm{i}$).
Interestingly the square of the Dirac operator is a topological Laplacian represented by a block diagonal matrix formed by two blocks: one acting on the nodes degree of freedom and reducing to the graph Laplacian and one acting on the link degree of freedom and reducing to the 1-up-Laplacian, i.e.
\bea
\mathbcal{L}=(\mathbcal{D})^2=\left(\begin{array}{cc} {\bf L}_{[0]}&0\\0&{\bf L}_{[1]}^{down}\end{array}\right).
\eea
In this work we want to use this Dirac operator to define a topological Dirac equation defined on a network.
Quantum wave equations including the Dirac equations and the Schr\"oedinger equations are typically defined on continuous space-time. When these equations are put on a lattice typically it is assumed that the wave function is defined on the nodes of the lattice (network). Here we assume instead that the wave function of a network is not only defined by a set of variables $\bm\phi$ defined on the nodes of the network but includes also   a set of variable $\bm\chi$ defined on the links of the network. Therefore we relax a bit the notion of point-like definition of the wave function and consider also degree of freedoms associated to links. Therefore we consider the wave function described by the topological spinor  
\bea
\bm\psi=\left(\begin{array}{c} \bm\phi\\ \bm\chi \end{array}\right),
\label{spinor}
\eea
with $\bm\phi$ indicating a $0$-cochain and $\bm\chi$ indicating a $1$-cochain associated to the network, i.e.
\bea
\bm \phi=\left(\begin{array}{c} \phi_1\\ \phi_2\\\vdots \\\phi_N\end{array}\right), \quad \bm \chi=\left(\begin{array}{c} \chi_{\ell_1}\\ \chi_{\ell_2}\\\vdots \\\chi_{\ell_M}\end{array}\right),
\eea
where a $0$-cochain  and a $1$-cochain indicate  functions defined on nodes $ \{1,2,\ldots N\}$ and links $\{\ell_1,\ell_2,\ldots, \ell_M\}$ respectively.
While in Sec.~5 we will touch on the problem of   modelling a discrete Lorentzian space-time, to start with  we consider that the network, although of arbitrary topology only captures the discrete space, while we keep time as a continuous variable.
The {\em topological Dirac equation } is defined as the wave-equation 
\bea
i \partial_t\bm\psi={\mathbcal{H}}\bm\psi.
\label{Topo_Dirac}
\eea
where the Hamiltonian of the topological Dirac equation is taken to be 
\bea
\mathbcal{H}=\mathbcal{D}+m_0\bm\beta.
\label{H}
\eea
Here $m_0\geq 0$ is a parameter of the dynamics we call the {\em mass} and the matrix $\bm\beta$ is a $(N+L)\times (N+L)$ block diagonal matrix of elements
\bea
\bm \beta=\left(\begin{array}{cc}{\bf I}_N & {\bf 0}\\ {\bf 0}& -{\bf I}_M\end{array}\right),
\label{beta0}
\eea
where ${\bf I}_P$ indicates the identity matrix of dimension $P$.
Interestingly we note that the anti-commutator between 
$\mathbcal{D}$ and $\bm\beta$ vanishes, i.e.
\bea
\{{\mathbcal D},\bm\beta\}={\bf 0},
\eea
and that \bea
\bm \beta^2={\bf I}_{N+M}.
\eea
Let us now consider the eigenstates at constant energy, satisfying 
\bea
E\bm\psi= \mathbcal{H}\bm\psi.
\label{E}
\eea
Let us write this equation explicitly by considering the part of the wave function $\bm\phi$ defined on nodes and the part of the wave function $\bm\chi$ defined on links. In this way we obtain, 
\bea
E\bm\phi =b{\bf B}\bm\chi+m_0\bm\phi,\nonumber \\
E\bm\chi =b^{\star}{\bf B}^{\dag} \bm\phi-m_0\bm\chi,
\eea
or equivalently 
\bea
(E-m_0)\bm\phi= b {\bf B}\bm\chi,\nonumber \\
(E+m_0)\bm\chi=  b^{\star}{\bf B}^{\dag} \bm\phi.
\label{Epm}
\eea
From this equation it can be shown that $\bm\phi$ should be an eigenvector of the graph Laplacian ${\bf L}_{[0]}$  and $\bm\chi$ should be an eigenvector of the 1-down-Laplacian ${\bf L}_{[1]}^{down}$. In fact with simple manipulations of Eqs. (\ref{Epm}) we obtain
\bea
(E-m_0)(E+m_0)\bm\phi={\bf B}{\bf B}^{\dag} \bm\phi={\bf L}_{[0]}\bm\phi,\\
(E-m_0)(E+m_0)\bm\chi={\bf B}^{\dag}{\bf B}\bm\chi={\bf L}_{[1]}^{down} \bm\chi.
\eea
By indicating with $\lambda$ the  generic eigenvalue of ${\bf B}$ and with $\lambda^{\star}$ the corresponding eigenvalue of ${\bf B}^{\dag}$ this equation implies the energy $E$ satisfy the relativistic dispersion relation
\bea
E^2=|\lambda|^2+m_0^2.
\label{dis}
\eea
This means that there are positive and negative energy eigenstates corresponding to energy values $E=\pm\sqrt{|\lambda|^2+m^2}$.
As long as the mass is strictly positive, i.e. $m_0>0$ the eigenfunctions associated to the positive and negative solutions are given by $\bm\psi_{\lambda,+}$ and $\bm\psi_{\lambda,-}$ respectively. These eigenstates are  given by 
\bea
{\bm \psi}_{\lambda,+}=\left(\frac{(|E|+m_0)}{2m_0}\right)^{1/2}\left(\begin{array}{c} {\bf w}_{\lambda}^{L}\\ \frac{b^{\star}\lambda^{\star}}{|E|+m_0}{\bf w}_{\lambda}^{R}\end{array}\right)
\\
{\bm\psi}_{\lambda,-}=\left(\frac{(|E|+m_0)}{2m_0}\right)^{1/2}\left(\begin{array}{c}  -\frac{b\lambda}{|E|+m_0}{\bf w}_{\lambda}^{L}\\ {\bf w}_{\lambda}^{R}\end{array}\right),
\eea
where ${\bf w}_{\lambda}^{R}$ and ${\bf w}_{\lambda}^{L}$ are the right and left eigenvectors of the incidence matrix ${\bf B}$ associated to the eigenvalue $\lambda$, and 
where we adopted the normalization condition 
\bea
\langle\bar{\bm \psi}_{\lambda,\pm}|{\bm \psi}_{\lambda,\pm}\rangle>=\langle{\bm \psi}_{\lambda,\pm}|\bm\beta|{\bm \psi}_{\lambda,\pm}\rangle>=\pm 1,
\eea
similar to the one used for the massive Dirac equation \cite{ryder1996quantum}.
In Figure $\ref{fig:1}$ and Figure $\ref{fig:2}$ we represent examples of  eigenstates of the topological Dirac equation corresponding to positive and negative energy values $E$  obtained starting from  real network datasets. As it is clear from the   figures and from the analytical expression of the eigenstates, these   eigenstates are strongly depending on the topology of the underlying network and on its spectral properties.
Let us now consider  the density of states associated to the dispersion relation Eq. (\ref{dis}). We observe that $\mu=|\lambda|^2$ indicates the eigenvalue of the graph Laplacian ${\bf L}_{[0]}$.  Indicating by $\rho(\mu)$ the density of positive eigenvalue of the graph Laplacian we can obtain the density $g(E)$ of positive energy eigenvalues $E>m_0$ from  the relation
\bea
g(E)dE=\rho(\mu)d\mu.
\eea
Using $E=\sqrt{\mu+m_0^2}$ we therefore obtain
\bea
g(E)=2E\rho(E^2-m_0^2).
\eea
In particular this implies that if the network has a spectral dimension $d_S$ \cite{rammal1983random,burioni2005random}, i.e. if $\rho(\mu)$ scales like
\bea
\rho(\mu)\propto \mu^{d_S/2-1}\quad \mbox{for}\  \mu\ll 1, 
\label{rhomuds}
\eea
then $g(E)$ scales as 
\bea
g(E)\propto E(E^2-m_0^2)^{d_S-2}\quad \mbox{for} \ 0<E^2-m_0^2\ll 1.\label{gEds}
\eea
Since the non-zero eigenvalues of the graph Laplacian ${\bf L}_{[0]}$ are the same eigenvalues of the 1-down Laplacian ${\bf L}_{[1]}^{down}$ (and with the same degeneracy) the eigenstates  at negative energy $E=-\sqrt{|\lambda|^2+m_0^2}$ have the same degeneracy of the states of positive energy $E=\sqrt{|\lambda|^2+m_0^2}$  as long as $|E|>m_0$. Therefore  we have that the density of negative energy eigenstates is given by $g(-E)$ as long as $|E|>m_0$. In Figure $\ref{fig:3}$ we display the cumulative density of eigenvalues $G(E)$ indicating the fraction of positive eigenvalues with energy  greater than $m_0$ and smaller than $E$ for the real networks investigated in Figure ${\ref{fig:1}}$ and Figure $\ref{fig:2}$. 
The density of states at energy $E=m_0$ and energy $E=-m_0$ deserves some particular attention. These energy states correspond to the eigenvalue $\mu=0$ of the graph Laplacian ${\bf L}_{[0]}$ and of  the 1-down Laplacian ${\bf L}_{[1]}^{down}$.
These two matrices have the same non-zero eigenvalues. In particular the positive eigenvalues coincide together with their degeneracy. However the matrix ${\bf L}_{[0]}$ is a $N\times N$ matrix and the matrix ${\bf L}_{[1]}$ is a $M\times M$ matrix therefore the zero eigenvalue has a different degeneracy for the two Laplacians. 
If we assume, without loss of generality that  the network is connected, then the number of links $M$ satisfies  $M\geq N-1$ where $M=N-1$ implies that the network is a tree and $L=N$ implies that the network is a ring (or chain with periodic boundary conditions).
An important result of spectral graph theory is that for a connected network the graph Laplacian ${\bf L}_{[0]}$ has a zero eigenvalue $\mu=0$ with multiplicity $1$ (corresponding to the fact that the network has a single connected component).
The ${\bf L}_{[1]}^{down}$ Laplacian of the same network will have $N-1$ non-zero eigenvalues while the multiplicity of the zero eigenvalue will be $M-(N-1)$. In other words  the multiplicity of the zero eigenvalue of the $1$-down-Laplacian ${\bf L}_{[1]}^{down}$ is equal to the number of independent cycles in the network.

\subsection{Topological Dirac equation on a   one dimensional chain}
The topological Dirac equation can be defined on a network of arbitrary topology, however it is instructive to investigate its properties on a $1$-dimensional chain.
For a $1$-dimensional chain of size $N$ with periodic boundary conditions we have that the number of links is equal to the number of nodes, (i.e. $M=N$) and that the eigenvalues of the graph Laplacian $\mu$ can be expressed as
\bea
\mu=|\lambda_x|^2=2\sin(k_x/2),
\eea
where  $k_x=2\pi \bar{n}_x/N$ indicates the wave-number and $0\leq \bar{n}_x\leq N-1$.
For $k_x\ll 1$ we can approximate the expression for $\mu$ with  
\bea
\mu=|\lambda_x|^2\simeq k_x^2.
\eea
The density of eigenvalues $\rho(\mu)\propto k_x^{-1/2}$ and the degeneracy of the eigenstates with positive energy $E=m_0$ and negative energy $E=-m_0$ is the same.
The eigenstates of the topological Dirac equation are formed by a $0$-cochain $\bm\phi$ and a $1$-cochain $\bm\chi$ that are proportional to the left and right eigenvector  of the incidence matrix ${\bf B}$ forming a $1$ dimensional Fourier basis with wave-number $k_x$. 
In this case the topological Dirac equation  is most similar to the  Dirac equation defined in one dimension  \cite{shen2012topological} with the most notable difference that the spinor is given a topological interpretation.
However the topological Dirac equation applied to higher dimensional lattices does not distinguish between the the discrete gradient (performed by the application of the incidence matrix ${\bf B}$) performed in different orthogonal directions. In this respect the topological Dirac equation differs significantly from the Dirac equation in $d=2$ and $d=3$ dimensions. In Sec.~3 we will  investigate how the topological Dirac equation can be modified to get closer to the Dirac equation in dimensions $d=2$ and $d=3$  taking into account the different directions  of  the links.
Before we explore this generalization, however, we devote the next section to treat the topological Dirac equation on simplicial complexes.


\section{Topological  Dirac equation of simplicial complexes}
Simplicial complexes are generalized network structures which  explicitly indicate the existence of higher-order interactions. Indeed simplicial complexes do not only include nodes and links but also include higher-order simplices such as triangles, tetrahedra and so on.
A $n$-dimensional simplex $\alpha$ is a set of $n+1$ nodes 
\bea
\alpha=[i_0,i_1,i_s,\ldots, i_n].
\label{simplex}
\eea
A simplicial complex $\mathcal{K}$ is a set of simplices closed under the inclusion of subsets. Therefore if a simplex $\alpha$ belongs to the simplicial complex, i.e. $\alpha\in \mathcal{K}$ any simplex $\alpha'\subset \alpha$ including a subset of the nodes of $\alpha$ also belongs to the simplicial complex, i.e. $\alpha'\in \mathcal{K}$. Here an in the following we indicate with $N_{[n]}$ the number of $n$-dimensional simplices of the considered simplicial complex. Therefore $N_{[0]}=N$ indicates the number of nodes of the simplicial complex and $N_{[1]}=M$ indicates the number of links of the simplicial complex.
Simplicial complexes describe discrete topological spaces and are  the fundamental objects studied by algebraic topology. Interestingly, when we associate a length to the links of the simplices, simplicial complexes can also be used to study arbitrary discrete geometries, and as such they have been extensively used in quantum gravity to describe discrete (or discretized) space-time.
Here we build on algebraic topology to formulate a higher-order version of the topological Dirac equation on simplicial complexes.

In algebraic topology \cite{hatcher}, simplices are also given an orientation, typically chosen to be induced by the node labels, so that for instance a link 
$[i,j]$ has positive orientation if $j>i$ and negative otherwise. 
Oriented $n$-dimensional  simplices can be chosen as the base of the space of $n$-chains $\mathcal{C}_n$ whose elements  are linear combinations of $n$-dimensional simplices with coefficients in $\mathbb{Z}$ (or $\mathbb{R}$).
The definition of $n$-chains allows us to explore the topology of the simplicial complex algebraically.
An important operator in algebraic topology is the {\em boundary operator} $\partial_n:\mathcal{C}_n\to \mathcal{C}_{n-1}$ that maps each $n$-simplex to the $(n-1)$-chain representing the $(n-1)$-dimensional simplices at its boundary.
The boundary operator $\partial_n$ indeed acts on any simplex $\alpha$ given by Eq. (\ref{simplex}) as 
\bea
\partial_n\alpha=\sum_{s=0}^n(-1)^s[i_0,i_1,\ldots i_{s-1},i_{s+1},\ldots i_n].
\eea
It follows that for $n=1$ every link is mapped to its two endnodes, i.e.
\bea
\partial_{1}[i,j]=[j]-[i],
\eea
and for $n=2$ every triangle is mapped to the three links at its boundary with the correct orientation, i.e. 
\bea
\partial_2[i,j,r]=[i,j]+[j,r]-[i,r].
\eea
Let us indicate with ${\bf B}_{[n]}$ the matrix that represents the $n$-boundary operator. We notice immediately that ${\bf B}_{[1]}={\bf B}$ indicates the incidence matrix of a network as defined in Eq. (\ref{B_link}) where this definition applies also to higher dimensional simplicial complexes.
The incidence matrices ${\bf B}_{[n]}$ have the notable property that their subsequent concatenation is null, i.e. for any $n>1$
\bea
{\bf B}_{[n-1]}{\bf B}_{[n]}={\bf 0}.
\label{boundary_null}
\eea
This  algebraic relation reflects  the important topological property that the boundary of the boundary is null.
The higher-order incidence matrices can be used to define the higher-order Laplacians ${\bf L}_{[n]}$. For $n=0$ the higher-order Laplacian reduces to the graph Laplacian ${\bf L}_{[0]}$ defined in Eq. (\ref{L0}), for $n>0$ the higher-order Laplacian is given by 
\bea
{\bf L}_{[n]}={\bf L}_{[n]}^{down}+{\bf L}_{[n]}^{up}
\eea 
with 
\bea
{\bf L}_{[n]}^{down}=[{\bf B}_{[n]}]^{\dag}{\bf B}_{[n]},\nonumber \\
{\bf L}_{[n]}^{up}={\bf B}_{[n+1]}[{\bf B}_{[n+1]}]^{\dag}. 
\eea
Interestingly ${\bf L}_{[n]}^{up},{\bf L}_{[n]}^{down}$ and ${\bf L}_{[n]}$ commute and therefore can be simultaneously diagonalized. Moreover thanks to the Hodge decomposition \cite{hatcher} any non-zero eigenvector of ${\bf L}_{[n]}$ is either a non-zero eigenvector of ${\bf L}_{[n]}^{up}$ or a non-zero eigenvector of ${\bf L}_{[n]}^{down}$.

Given this important insight for algebraic topology we can generalize the topological Dirac equation to higher-order topologies.
In particular we can define the $n$-order Dirac operator with $n\geq 1$ as an operator represented by a $(N_{[n-1]}+N_{[n]})$ square matrix given by 
\bea
\mathbcal{D}_{[n]}=\left(\begin{array}{cc}{\bf 0}& b_{[n]}{\bf B}_{[n]}\\ b_{[n]}^{\star}{\bf B}_{[n]}^{\dag}& {\bf 0}\end{array}\right),
\eea
where $b_{[n]}\in \mathbb{C}$ has absolute value $|b_{[n]}|=1$ and where $b_{[n]}^{\star}$ is its complex conjugate. 
Therefore we have that for $n=1$ the $n$-order Dirac operator reduces to the Dirac operator $\mathbcal{D}$ defined in the previous paragraph.
The square of the $n$-order Dirac operator leads to the Laplacian operator acting on $(n-1)$-dimensional  simplices and $n$-dimensional simplices and describing diffusion among $(n-1)$-simplices through $n$-simplices and diffusion among $n$ simplices through $(n-1)$-simplices, i.e.  
\bea
\mathbcal{L}_{[n]}=(\mathbcal{D}_{[n]})^2=\left(\begin{array}{cc} {\bf L}_{[n-1]}^{up}&{\bf 0}\\ {\bf 0}&{\bf L}_{[n]}^{down}\end{array}\right).
\eea
Using the $n$-order Dirac operator we can define the $n$-order topological Dirac equation of a simplicial complex. This equation is defined on a topological spinor $\bm\psi_{[n]}$ including a $(n-1)$-cochain $\bm\chi_{[n-1]}$ and a $n$-cochain $\bm\chi_{[n]}$, i.e. the topological spinor $\bm\psi_{[n]}$ is a $(N_{[n-1]}+N_{[n]})$-dimensional column vector given by 
\bea
\bm\psi_{[n]}=\left(\begin{array}{c} \bm\chi_{[n-1]}\\\bm{\chi}_{[n]} \end{array}\right).
\label{spinor_simplex}
\eea
According to the definition used in this section $\bm\chi_{[0]}$ is the $0$-cochain that can be identified with the $0$-cochain indicated in the previous section by $\bm\phi$. Similarly  $\bm\chi_{[1]}$ can be identified with $\bm\chi$ used in the previous section.
The $n$-order topological Dirac equation on simplicial complex, describes the evolution of the $n$-order topological spinor $\bm\psi_{[n]}$ according to the equation
\bea
i \partial_t\bm\psi_{[n]}=\mathbcal{H}_{[n]}\bm\psi_{[n]}.
\label{Topo_Dirac_simplex}
\eea
where the Hamiltonian $\mathbcal{H}_{[n]}$, given by 
\bea
\mathbcal{H}_{[n]}=\mathbcal{D}_{[n]}+m_0\bm\beta_{[n]}
\eea
is defined in terms of the $n$-order Dirac operator $\mathbcal{D}_{[n]}$ and the matrix $\bm\beta_{[n]}$ given by 
\bea
\bm\beta_{[n]}=\left(\begin{array}{cc}{\bf I}_{N_{[n-1]}} & {\bf 0}\\ {\bf 0}& -{\bf I}_{N_{[n]}}\end{array}\right).
\eea
By indicating with $\lambda_n$ the generic eigenvalue of the boundary operator ${\bf B}_{[n]}$ and proceeding in analizing the eigenstate of the Hamiltonian using  steps similar to the ones used in the previous paragraph, one can easily see that the energy $E_{[n]}$ associated to the eigenstates of ${\bf H}_{[n]}$ satisfies the dispersion relation
\bea
E^2_{[n]}=|\lambda_n|^2+m_0^2.
\eea
As long as the mass $m_0$ is strictly positive the eigenstates associated the the positive energy states ($\bm\psi_{\lambda,+}$) and the ones associated with the negative energy states ($\bm\psi_{\lambda,-}$) can be normalized as the eigenstates of the topological Dirac equation on  networks and they are given by 
\bea
{\bm \psi}_{[n],\lambda,+}=\left(\frac{(|E|+m_0)}{2m_0}\right)^{1/2}\left(\begin{array}{c} {\bf w}_{[n]\lambda}^{L}\\ \frac{b^{\star}_{[n]}\lambda^{\star}}{|E|+m_0}{\bf w}_{[n]\lambda}^{R}\end{array}\right)
\\
{\bm\psi}_{[n],\lambda,-}=\left(\frac{(|E|+m_0)}{2m_0}\right)^{1/2}\left(\begin{array}{c}  -\frac{b_{[n]}\lambda}{|E|+m_0}{\bf w}_{[n]\lambda}^{L}\\ {\bf w}_{[n]\lambda}^{R}\end{array}\right)
\eea
where ${\bf w}_{[n]\lambda}^{R}$ and ${\bf w}_{[n]\lambda}^{L}$ are the right and left eigenvectors of the boundary operator ${\bf B}_{[n]}$ associated to the eigenvalue $\lambda$.
Interestingly, it is also possible to consider a higher-order topological Dirac equation defined on a spinor including all possible $n$-cochains defined on the simplicial complex.
Therefore, if the simplicial complex is $d$ dimensional, i.e. if the simplicial complex includes simplices of dimensions up to $n=d$, we define the topological spinor
\bea
\bm \Psi_{[d]}=\left(\begin{array}{c} \bm\chi_{[0]}\\\bm{\chi}_{[1]}\\\vdots\\\bm\chi_{[d-1]}\\\bm\chi_{[d]} \end{array}\right).
\eea
and we consider the higher-order topological Dirac equation given by 
\bea
i \partial_t\bm\Psi_{[d]}=\bar\mathbcal{H}_{[d]}\bm\Psi_{[d]}.
\label{Topo_Dirac_simplex_h}
\eea
with Hamiltonian $\tilde{\mathbcal{H}}_{[d]}$ given by
\bea
\tilde{\mathbcal{H}}_{[d]}=\sum_{n=0}^{d}\left(\mathbcal{D}_{[n]}+m_0\bm\beta_{[n]}\right)
\eea
where now we have lifted the dimension of $\mathbcal{D}_{[n]}$ and the dimension of $\bm\beta_{[n]}$ to the dimension of the spinor $\bm\Psi_{[d]}$. This leads to the matrix representation for $\mathbcal{D}_{[n]}$ and $\bm\beta_{[n]}$ given by 
\bea
\hspace*{-10mm}\mathbcal{D}_{[n]}=\left(\begin{array}{cccccc}{\bf 0}_{N_{[0]}\times N_{[0]}}&\ldots &{\bf 0}_{N_{[0]}\times N_{[n-1]}}& {\bf 0}_{N_{[0]}\times N_{[n]}}&{\bf 0}_{N_{[n-1]}\times N_{[n+1]}}\ldots& {\bf 0}_{N_{[0]}\times N_{[d]}}\\\vdots & \vdots &\vdots &\vdots & \vdots & \vdots \\ {\bf 0}_{N_{[n-1]}\times N_{[0]}}&\ldots &{\bf 0}_{N_{[n-1]}\times N_{[n-1]}}& b_{[n]}{\bf B}_{[n]}&{\bf 0}_{N_{[k-1]}\times N_{[n+1]}}\ldots& {\bf 0}_{N_{[k-1]}\times N_{[d]}}\\{\bf 0}_{N_{[n]}\times N_{[0]}}&\ldots &b^{\star}_{[n]}{\bf B}_{[n]}^{\dag}& {\bf 0}_{N_{[n]}\times N_{[n]}} & {\bf 0}_{N_{[n]}\times N_{[n+1]}}\ldots & {\bf 0}_{N_{[n]}\times N_{[d]}}\\\vdots& \vdots&\vdots&\vdots&\vdots&\vdots\\
{\bf 0}_{N_{[d]}\times N_{[0]}}&\ldots &{\bf 0}_{N_{[d]}\times N_{[n-1]}}& {\bf 0}_{N_{[d]}\times N_{[n]}}&{\bf 0}_{N_{[d]}\times N_{[n+1]}}\ldots& {\bf 0}_{N_{[d]}\times N_{[d]}}\\\end{array}\right),\nonumber\\
\hspace*{-10mm}\bm\beta_{[n]}=\left(\begin{array}{cccccc}{\bf 0}_{N_{[0]}\times N_{[0]}}&\ldots &{\bf 0}_{N_{[0]}\times N_{[n-1]}}& {\bf 0}_{N_{[0]}\times N_{[n]}}&{\bf 0}_{N_{[n-1]}\times N_{[n+1]}}\ldots& {\bf 0}_{N_{[0]}\times N_{[d]}}\\\vdots & \vdots &\vdots &\vdots & \vdots & \vdots \\ {\bf 0}_{N_{[n-1]}\times N_{[0]}}&\ldots &{\bf I}_{N_{[n-1]}}&{\bf 0}_{N_{[n-1]}\times N_{[n]}}&{\bf 0}_{N_{[n-1]}\times N_{[n+1]}}\ldots& {\bf 0}_{N_{[n-1]}\times N_{[d]}}\\{\bf 0}_{N_{[n]}\times N_{[0]}}&\ldots &{\bf 0}_{N_{[n]}\times N_{[n-1]}}&-{\bf I}_{N_{[n]}} & {\bf 0}_{N_{[n]}\times N_{[n+1]}}\ldots & {\bf 0}_{N_{[n]}\times N_{[d]}}\\\vdots& \vdots&\vdots&\vdots&\vdots&\vdots\\
{\bf 0}_{N_{[d]}\times N_{[0]}}&\ldots &{\bf 0}_{N_{[d]}\times N_{[n-1]}}& {\bf 0}_{N_{[d]}\times N_{[n]}}&{\bf 0}_{N_{[d]}\times N_{[n+1]}}\ldots& {\bf 0}_{N_{[d]}\times N_{[d]}}\\\end{array}\right).\nonumber
\eea
Note that from Eq. (\ref{boundary_null}) it follows that the $n$-order Dirac operators obey
\bea
\mathbcal{D}_{[n]}\mathbcal{D}_{[n']}={\bf 0},
\eea
for any $n\neq n'$.
Therefore  we have 
\bea
\left(\sum_{n=0}\mathbcal{D}_{[n]}\right)^2=\mathbcal{L},
\eea
with $\mathbcal{L}$ being the block diagonal matrix with diagonal blocks given by the $n$-order Laplacian, i.e.
\bea
\hspace*{-10mm}\mathbcal{L}_=\left(\begin{array}{cccccc}{\bf L}_{[0]}&\ldots &{\bf 0}_{N_{[0]}\times N_{[n-1]}}& {\bf 0}_{N_{[0]}\times N_{[n]}}&{\bf 0}_{N_{[n-1]}\times N_{[n+1]}}\ldots& {\bf 0}_{N_{[0]}\times N_{[d]}}\\\vdots & \vdots &\vdots &\vdots & \vdots & \vdots \\ {\bf 0}_{N_{[n-1]}\times N_{[0]}}&\ldots &{\bf L}_{[n-1]}&{\bf 0}_{N_{[n-1]}\times N_{[n]}}&{\bf 0}_{N_{[n-1]}\times N_{[n+1]}}\ldots& {\bf 0}_{N_{[n-1]}\times N_{[d]}}\\{\bf 0}_{N_{[n]}\times N_{[0]}}&\ldots &{\bf 0}_{N_{[n]}\times N_{[n-1]}}&{\bf L}_{[n]} & {\bf 0}_{N_{[n]}\times N_{[n+1]}}\ldots & {\bf 0}_{N_{[n]}\times N_{[d]}}\\\vdots& \vdots&\vdots&\vdots&\vdots&\vdots\\
{\bf 0}_{N_{[d]}\times N_{[0]}}&\ldots &{\bf 0}_{N_{[d]}\times N_{[n-1]}}& {\bf 0}_{N_{[d]}\times N_{[n]}}&{\bf 0}_{N_{[d]}\times N_{[n+1]}}\ldots& {\bf L}_{[d]}\\\end{array}\right).\nonumber
\eea
Interestingly this latter formulation of the higher-order Dirac equation on simplicial complexes admits eigenstates that are exclusively localized on $(n-1)$-dimensional and $n$-dimensional simplices for $0<n\leq d$ where $d$ is the dimension of the simplicial complex. 
To be concrete we can consider this higher-order topological Dirac equation on a simplicial complex of dimension $d=3$.
In this case the Hamiltonian $\tilde{\mathbcal{H}}_{[3]}$ reads 
\bea
\hspace*{-10mm}\tilde{\mathbcal{H}}_{[3]}=\left(\begin{array}{cccc}m_0{\bf I}_{N_{[0]}}&b_{[1]}{\bf B}_{[1]} &{\bf 0}_{N_{[0]}\times N_{[2]}}& {\bf 0}_{N_{[0]}\times N_{[3]}}\\ b_{[1]}^{\star}{\bf B}_{[1]}^{\dag}&{\bf 0}_{N_{[1]}\times N_{[1]}}&b_{[2]}{\bf B}_{[2]}&{\bf 0}_{N_{[1]}\times N_{[3]}}\\
{\bf 0}_{N_{[2]}\times N_{[0]}}&b^{\star}_{[2]}{\bf B}_{[2]}^{\dag}&{\bf 0}_{N_{[2]}\times N_{[2]}}&b_{[3]}{\bf B}_{[3]}\\
{\bf 0}_{N_{[3]}\times N_{[0]}}&{\bf 0}_{N_{[3]}\times N_{[1]}}&b^{\star}_{[3]}{\bf B}_{[3]}^{\dag}&-m_0{\bf I}_{N_{[3]}}\end{array}\right).
\eea
The eigenstate $\bm\Psi_{[3]}=(\bm\chi_{[0]},\bm\chi_{[1]},\bm\chi_{[2]},\bm\chi_{[3]})^{\top}$ corresponding to the energy $E$ satisfies the system of equations
\bea
(E-m_0)\bm\chi_{[0]}=b_{[1]}{\bf B}_{[1]}\bm\chi_{[1]}\nonumber \\
E\bm\chi_{[1]}=b_{[1]}^{\star}{\bf B}_{[1]}^{\dag}\bm\chi_{[0]}+b_{[2]}{\bf B}_{[2]}\bm\chi_{[2]}\nonumber \\
E\bm\chi_{[2]}=b^{\star}_{[2]}{\bf B}_{[2]}^{\dag}\bm\chi_{[1]}+b_{[3]}{\bf B}_{[3]}\bm\chi_{[3]}\nonumber \\
(E+m_0)\bm\chi_{[3]}=b_{[3]}^{\star}{\bf B}_{[3]}^{\dag}\bm\chi_{[2]}
\eea
This system of equations can be easily generalized to arbitrary dimension $d$. By investigating this equation it is possible to see that this system of equation admits $d$ positive and $d$ negative energy bands $E_{[n]}$ with the $n$th (positive and negative) energy bands depending only on the  eigenvalues of the $n$-incidence matrices $\lambda_n$ and the mass $m_0$. We have in particular that the energy dispersion relation of the different bands is 
\bea
E_{[1]}=\frac{m_0}{2}\pm \sqrt{\left(\frac{m_0}{2}\right)^2+|\lambda_1|^2,}\nonumber\\
E_{[n]}=\pm |\lambda_n|\ \mbox{for}\ 1<n<d,\nonumber\\
E_{[d]}=-\frac{m_0}{2}\pm \sqrt{\left(\frac{m_0}{2}\right)^2+|\lambda_d|^2}
\eea
where for small $|\lambda_1|\ll m_0 $ and small $|\lambda_d|\ll m_0$ we have
\bea
E_{[1]}=\frac{1}{2}(m_0\pm m_0)\pm \frac{|\lambda_1|^2}{m_0},
\nonumber\\
E_{[d]}=\frac{1}{2}(-m_0\pm m_0)\pm \frac{|\lambda_d|^2}{m_0}.
\eea
\begin{figure}[!htbp]
\begin{center}
\includegraphics[width=0.98\textwidth]{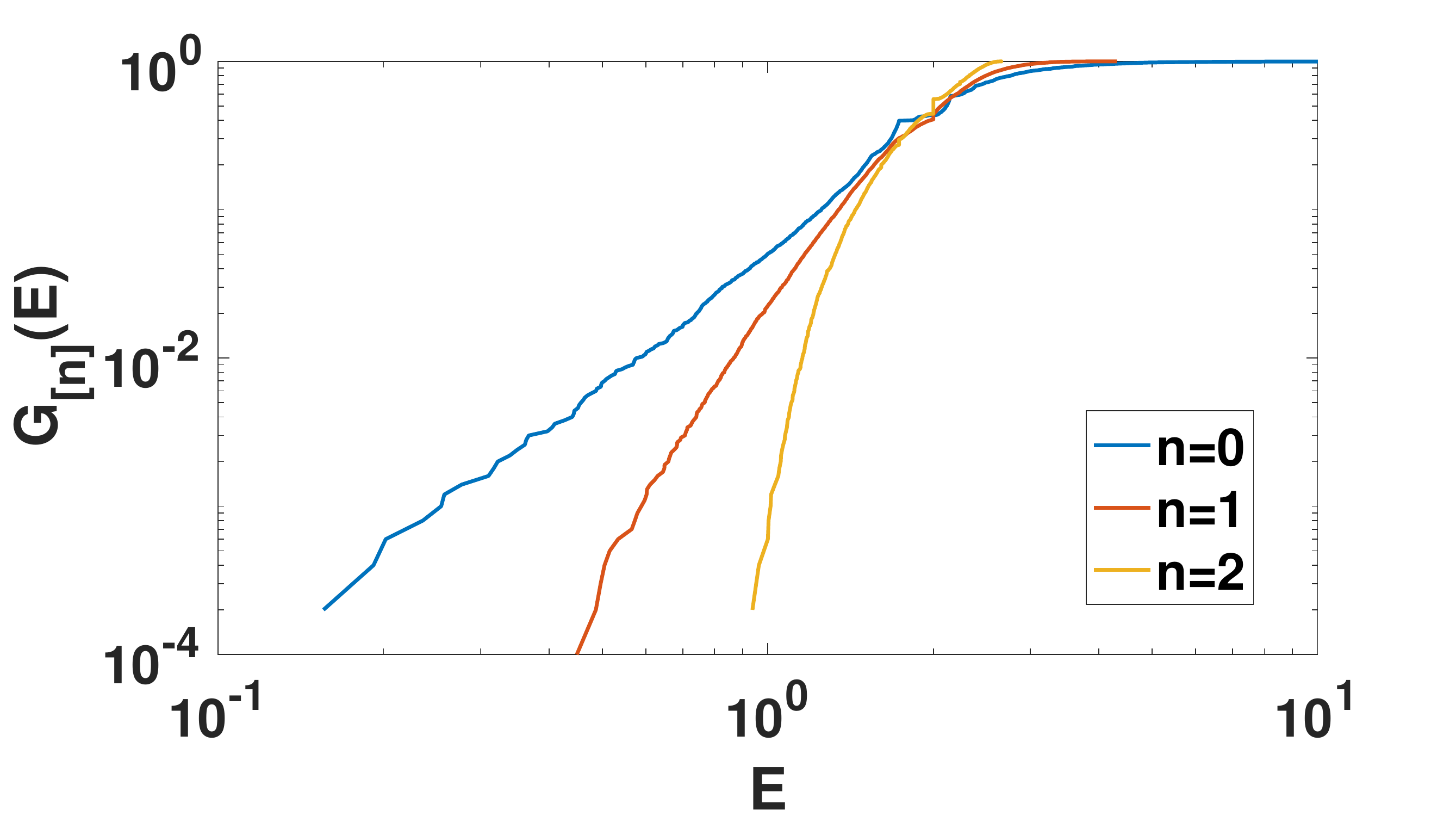}
\caption{{\bf  Different energy bands of the topological Dirac equation on simplicial complexes.} The cumulative density of eigenvalues $G_{[n]}(E)$ of the $n$th energy band of the topological Dirac equation on a $d=3$ dimensional simplicial complex is plotted for  $n\in \{0,1,2\}$ and mass $m_0=0$. Here the simplicial complex is chosen to be a single realization of the  the ``Network Geometry with Flavor" \cite{bianconi2016network,bianconi2017emergent} for emergent hyperbolic geometry with dimension $d=3$, flavor $s=-1$ and null inverse temperature. }
\label{fig:4}
\end{center}
\end{figure}
Interestingly for $m_0=0$ all the energy bands have relativistic dispersion relation 
\bea
E_{[n]}=\pm|\lambda_n|.
\eea
For each band, the  density of states with $|E_{[n]}|>m_0$ can be calculated using the density of states $\rho_{[n-1]}(\mu)$ of the non-zero eigenvalues of the $(n-1)$-up Laplacian ${\bf L}_{[n-1]}^{up}$ using 
\bea
g_{[n]}(E)dE=\rho_{[n-1]}(\mu)d\mu.
\eea
In particular for the case in which the mass is null, $m_0=0$, we have
\bea
g_{[n]}(E)=2E\rho_{[n-1]}(E^2) 
\eea
for all $0<n\leq d$.
It has been recently pointed out that geometrical simplicial complexes can admit differebr higher-order  spectral dimensions $d_S^{[n]}$ \cite{reitz2020higher,torres2020simplicial} determining the scaling of $\rho_{[n]}(\mu)$ for $\mu\ll1$,  which obeys 
\bea
\rho_{[n]}(\mu)\propto \mu^{d_S^{[n]}/2-1}\quad \mbox{for}\  \mu\ll 1. 
\eea
In this case we obtain that the higher-order Dirac equation admits $d$ positive and $d$ negative energy bands with density of states $g_{[n]}$ obeying
\bea
g_{[n]}(E)\propto (E)^{d_S^{[n-1]}-1}\quad \mbox{for} \ m_0=0, 0<E\ll 1.
\eea
In Figure $\ref{fig:4}$ we show the multiband spectrum of  the simplicial complex model of emergent hyperbolic geometry called "Network Geometry with Flavor" \cite{bianconi2016network,bianconi2017emergent} showing that the different bands can be associated to different spectral dimensions.
The eigenstates associated to the $n$-th energy band are localized on $(n-1)$ and $n$-dimensional cochains. Possibly by introducing opportune interaction terms one could induce transitions among states in different bands and even  hybridizations of these different eigenstates.

We finish this section by noticing that  the topological Dirac equation can be also  extended to treat hypegraphs in which hyperedges are directed, i.e. the hyperedges  have input and output nodes such as in chemical reaction networks. Indeed for such hypegraphs the incidence matrix has been recently defined in Ref. \cite{jost2019hypergraph} which can easily allow  the generalization of the topological Dirac operator and Dirac equation to hypegraphs. This is a very interesting extension of the topological Dirac equation, however due to space limitation we omit this discussion in the present paper.

\section{Directional topological Dirac equations on lattices}

\subsection{Directional  topological Dirac equation in  $d=2$-dimensional lattices}
In general network topologies describing discrete spaces with non trivial local fluctuations of the curvature it might be non trivial to define the equivalent to the orthogonal directions of lattices. For this reason the topological Dirac equation introduced in Sec.~1  only includes the $\bm \beta$ matrix but not the equivalent of all the $\bm\gamma$ matrices present in the $d=3$ dimensional Dirac equation. However  for $2$-dimensional and $3$-dimensional lattices the orthogonal spatial directions are naturally defined. Therefore  we can  explore whether it is possible to define a variation of the topological Dirac equation which distinguishes between the different spatial directions.

\begin{figure}[!htbp]
\begin{center}
\includegraphics[width=0.98\textwidth]{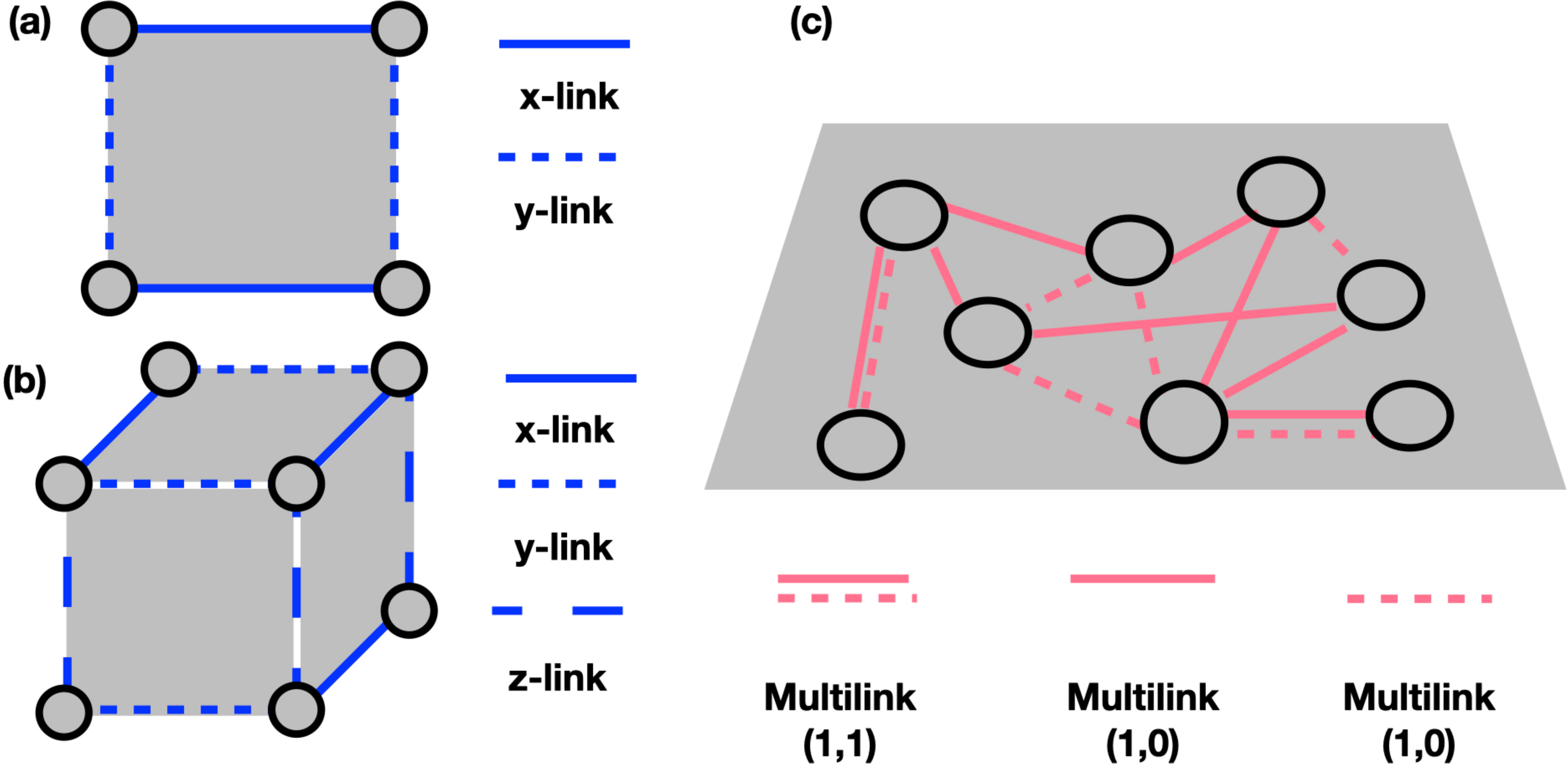}
\caption{{\bf Different types of links on lattices and multiplex networks can be treated differently by the directional topological Dirac operator.} On a unitary cell of a $2$-dimensional lattice (panel a) the topological Dirac equation distinguishes between $x$-links and $ y$-links. On a unitary cell of a $3$-dimensional lattice (panel b) links of type $x$, $y$, $z$ are distinguished. On a multiplex network (panel c) the directional topological Dirac equation distinguishes between different types of multilinks $\vec{m}\in \{(1,0),(0,1),(1,1)\}$.}
\label{fig:5}
\end{center}
\end{figure}

To this end we consider a square portion of a square lattice with sides of length $\sqrt{N}$ with periodic boundary condition (a torus) and we classify links of the $d=2$ square lattice according to their direction (see Figure $\ref{fig:5}$). 
For instance we consider $x$-type links the links connecting adjacent  nodes along the $x$-direction and  $y$-type links  connecting adjacent nodes along the $y$-direction. Therefore a $x$-type link will connect a node $i$ of coordinates ${\bf r}_i=(x_i,y_i)$ to the nodes $j$ of coordinates ${\bf r}_j={\bf r_i}\pm {\bf e}_x$  modulo periodic boundary conditions, where ${\bf e}_x=(1,0)$.
Similarly a $y$-type link  will connect a node $i$ of coordinates ${\bf r}_i=(x_i,y_i)$ to the nodes $j$ of coordinates ${\bf r}_j={\bf r_i}\pm {\bf e}_y$  modulo periodic boundary conditions, where ${\bf e}_y=(0,1)$.

Given this classification of different types of links, distinguished according to   their direction,  the incidence matrix ${\bf B}$ of a $d=2$ dimensional lattice can be written as 
\bea
{\bf B}={\bf B}_{(x)}+{\bf B}_{(y)},
\eea
  where  ${\bf B}_{(w)}$ with $w\in \{x,y\}$ describes the incidence matrix between nodes and links of type $w$, i.e. 
\bea
[{\bf B}_{(w)}]_{i\ell}=\left\{\begin{array}{cccc}1 & \mbox{if}\   \ell=[j,i], &\mbox{and } \ell\  \mbox{is a $w$-type link }\\ -1 & \mbox{if} \ \ell=[i,j],  &\mbox{and } \ell\  \mbox{is a $w$-type link }\\  0 & \  \mbox{otherwise}. &\end{array}\right.
\label{Bdir}
\eea
Introducing this decomposition of the incidence matrix in contributions coming from links of orthogonal directions opens the possibility to decompose also the Dirac operation accordingly and to introduce some additional rotation in the topological space of nodes and links.
To this end, for lattices of dimension $d=2$ we can define the directional Dirac operator $\bar\mathbcal{D}$ as 
\bea
\bar\mathbcal{{D}}=\sum_{w\in (x,y)}\mathbcal{D}_{(w)},
\label{2dD}
\eea
with $\mathbcal{D}_{(w)}$ describing the contribution of $w$-type links to the directional Dirac operator. We can proceed by considering rotations due to the different directions of the links. Therefore, inspired by the standard Dirac equation in dimension $d=2$ \cite{shen2012topological} we define $\mathbcal{D}_{(x)}$ and $\mathbcal{D}_{(y)}$ as
\bea \mathbcal{D}_{(x)}=\left(\begin{array}{cc} 0& {\bf B}_{(x)}\\ {[{\bf B}_{(x)}]}^{\dag} &0\end{array}\right),
\quad \mathbcal{D}_{(y)}=\left(\begin{array}{cc}0& \mathrm{i}{\bf B}_{(y)}\\-\mathrm{i}{[{\bf B}_{(y)}]}^{\dag}&0\end{array}\right).
\eea
With this choice of the directional Dirac operator, we consider always the topological spinor $\bm\psi=(\bm\phi,\bm\chi)^{\top}$ of dimension $N+M$ defined on Eq. (\ref{spinor}) and we consider a modified topological Dirac equation given by 
\bea
i \partial_t\bm\psi=\bar{\mathbcal{H}}\bm\psi,
\eea
with  Hamiltonian $\bar{\mathbcal{H}}$  taken to be 
\bea
\bar\mathcal{H}=\bar\mathbcal{{D}}+m_0\bm\beta
\eea
where $\bar\mathbcal{{D}}$ is the directional Dirac operator in two dimension given by Eq. (\ref{2dD}) and the matrix $\bm\beta$ is given by Eq. (\ref{beta0}).
Note however a very notable consequence of this definition of directional topological Dirac equation aimed at treating differently orthogonal space directions: while the anti-commutator between $\mathbcal{D}_{(w)}$ and the matrix $\bm\beta$ is zeros for both $w\in\{x,y\}$, i.e.
\bea
\{\mathbcal{D}_{(w)},\bm\beta\}={\bf 0}
\eea
 the anti-commutator between $\mathbcal{D}_{(x)}$ and $\mathbcal{D}_{(y)}$ is given by the  not zero matrix
 \bea
 \{\mathbcal{D}_{(x)},\mathbcal{D}_{(y)}\}=\left(\begin{array}{cc} {\bf 0}& {\bf 0}\\ {\bf 0}&\mathrm{i}\left([{\bf B}_{(x)}]^{\dag}{\bf B}_{(y)}-{[{\bf B}_{(y)}]^{\dag}}{{\bf B}_{(x)}} \right)\end{array}\right).
 \label{Ddir2d}
 \eea
 This is a notable difference with respect to the Dirac equation where the corresponding anti-commutator vanishes as the derivative respect to $x$ and to $y$ commute.
 Despite this difference it can be shown that the directional Dirac operator has eigenfunctions that satisfy the relativistic dispersion relation 
 \bea
 E=|{\bm\lambda}|^2+m^2,
 \eea
 where $\bm\lambda=(\lambda_x,\lambda_y)$ with $\lambda_x$ and $\lambda_y$ representing the eigenvalues of the directional incidence matrices ${\bf B}_{(x)}$ and ${\bf B}_{(y)}$ respectively.
 In particular we have that $\lambda_x$ and $\lambda_y$ can be expressed in terms of the wave-number ${\bf k}=(k_x,k_y)$ as
 \bea
 |\lambda_x|=2\sin(k_x/2),\nonumber\\
 |\lambda_y|=2\sin(k_y/2),
 \eea 
 where $k_x=2\pi \bar{n}_x/\sqrt{N}$ and $k_y=2\pi \bar{n}_y/\sqrt{N}$ with $\bar{n}_x,\bar{n}_y$ integers between zero and $N^{1/2}-1$.
 In order to show this results we can decompose the link component $\bm\chi$ of the topological spinor in two blocks,  $\bm\chi=(\bm\chi_{(x)}, \bm\chi_{(y)})^{\top}$ where $\bm\chi_{(w)}$ indicates the wave function calculated on links of type $w$.
 The wave function at constant energy $E$ then satisfies Eq. (\ref{E}) which can be rewritten as 
 \bea
 E\bm\phi={\bf B}_{(x)}\bm\chi_{(x)}+\mathrm{i}{\bf B}_{(y)}\bm\chi_{(y)}+m_0\bm\phi,\nonumber\\
  E\bm\chi_{(x)}=[{\bf B}_{(x)}]^{\dag}\bm\phi-m_0\bm\chi_{(x)}, \nonumber\\
    E\bm\chi_{(y)}=-\mathrm{i}[{\bf B}_{(y)}]^{\dag}\bm\phi-m_0\bm\chi_{(y)}.
 \eea
Simple algebraic operations show that $\bm\phi$ should be an eigenvector of the graph Laplacian, 
 \bea
 (E-m_0)(E+m_0)\bm \phi={\bf L}_{[0]}\bm\phi
 \eea
 with the graph Laplacian  ${\bf L}_{[0]}$ given by
 \bea
 {\bf L}_{[0]}={\bf L}_{(x)}+{\bf L}_{(y)},
 \eea
where
 \bea
{\bf L}_{(x)}= {\bf B}_{(x)}[{\bf B}_{(x)}]^{\dag},\quad {\bf L}_{(y)}={\bf B}_{(y)}[{\bf B}_{(y)}]^{\dag}.
 \eea
We note that ${\bf L}_{(x)}$ and ${\bf L}_{(y)}$ commute, i.e.
 \bea
 [{\bf L}_{(x)},{\bf L}_{(y)}]={\bf 0},
 \eea
therefore  the eigenstates of the directional Dirac equation have a dispersion relation 
 \bea
 E^2=m_0^2+\mu
 \eea
 where $\mu$ is the generic eigenvalue of the graph Laplacian ${\bf L}_{[0]}$ that can be written as 
 \bea
 \mu=|\lambda_x|^2+|\lambda_y|^2.
 \eea
 It follows that for $|{\bf k}|\ll1$, i.e. $\mu\ll1$ the density of eigenvalues $\rho(\mu)$ of the graph Laplacian follows Eq. (\ref{rhomuds}) and the density $g(E)$ of eigenstates of positive energy $E$ follows Eq. (\ref{gEds}) with $d_S=d=2$.
The eigenstates of the directional topological Dirac equation have the $\bm\phi$ component proportional to the eigenvectors of the graph Laplacian forming a $d=2$-dimensional Fourier basis with associated wave-number ${\bf k}=(k_x,k_y)$. In particular we have that the $\bm\phi=(\phi_1,\phi_2,\ldots, \phi_i\ldots, \phi_N)^{\top}$ component of the topological spinor at energy $E=\pm\sqrt{\lambda_x^2+\lambda_y^2+m_0^2}$ has elements
\bea
\phi_i=\mathcal{N} e^{\mathrm{i} {\bf k}\cdot {\bf r}_i},
\eea
where ${\bf k}=(k_x,k_y)$ and 
where $\mathcal{N}$ is a normalization constant, 
while the components $\bm\chi_{(x)}$ and $\bm\chi_{(y)}$ satisfy
\bea
\bm\chi_{(x)}&=&\frac{1}{(E+m_0)}{\bf B}_{(x)}^{\dag}\bm\phi,\nonumber \\
\bm\chi_{(y)}&=&\frac{1}{(E+m_0)}{\bf B}_{(y)}^{\dag}\bm\phi.
\eea 
Here we note that since  ${\bf L}_{[0]}$ commutes with ${\bf L}_{(w)}$ the component $\bm\chi$  of the eigenstates of the directional topological Dirac equation is also  eigenvector of the ${\bf L}_{[1],(x)}^{down}$ and the ${\bf L}_{[1],(y)}^{down}$ Laplacians given by 
\bea
{\bf L}_{[1],(x)}^{down}={\bf B}_{(x)}^{\dag}{\bf B}_{(x)},\quad{\bf L}_{[1],(y)}^{down}={\bf B}_{(y)}^{\dag}{\bf B}_{(y)}.
\eea

 \subsection{Directional topological Dirac equation in $d=3$ dimensional lattices}

 In this section we consider a cubic portion of a  three dimensional lattice with sides of length $N^{1/3}$ and periodic boundary conditions.  We distinguish between links of type $x,y$ and $z$ according to their direction (see Figure $\ref{fig:5}$). In particular two nodes $i$ and $j$ of coordinates ${\bf r}_i=(x_i,y_i,z_i)$ and ${\bf r}_{j}=(x_j,y_j,z_j)$ are connected by a $w$-type link if and only if ${\bf r}_j={\bf r}_i\pm{\bf e}_{(w)}$ modulo periodic boundary conditions, where ${\bf e}_{(x)}=(1,0,0)$, ${\bf e}_{(y)}=(0,1,0)$, and ${\bf e}_{(z)}=(0,0,1)$.
The incidence matrix ${\bf B}$ of the three dimensional lattice can be written as 
 \bea
 {\bf B}={\bf B}_{(x)}+{\bf B}_{(y)}+{\bf B}_{(z)},
 \eea
where ${\bf B}_{(w)}$ indicates the directional incidence matrix between $w$-type links and nodes that admits the same definition as in two dimensions (Eq.(\ref{Bdir})).

It turns out that  for lattices in dimension $d=3$ the topological spinor of dimension $N+M$ is not sufficient to distinguish between the three orthogonal spatial directions. Indeed, if we want to define the directional topological Dirac operator that distinguishes between the three different types of links we need to consider the topological spinor $\tilde{\bm\Psi}$  of dimension $2(N+M)$ given by 
\bea
\tilde{\bm\Psi}=\left(\begin{array}{c} \bm\Phi\\ \bm X \end{array}\right),
\label{spinord3}
\eea
where $\bm\Phi$ is a vector of two $0$-cochains $\bm\phi^{(r)}$ with $r\in \{1,2\}$ defined on the nodes of the network and $\bf X$ is a vector of two $1$-cochains $\bm \chi^{(r)}$ with $r\in \{1,2\}$ defined on the links of the network, i.e.
\bea
\bm\Phi=\left(\begin{array}{c} \bm\phi^{(1)}\\\bm\phi^{(2)} \end{array}\right),\quad \bm X=\left(\begin{array}{c}  \bm\chi^{(1)}\\ \bm \chi^{(2)} \end{array}\right).
\eea
Having defined the directional incidence matrix and the topological spinor, the natural next step is to define a directional Dirac operator given by  
\bea
\bar\mathbcal{D}=\sum_{w\in (x,y,z)}\mathbcal{D}_{(w)},
\label{3dD}
\eea
where the $w$-directional Dirac operators $\mathbcal{D}_{(w)}$ define independent rotations of the topological spinor.
The $w$-directional Dirac operator $\mathbcal{D}_{(w)}$ has a block structure given by 
\bea
\mathbcal{D}_{(w)}=\left(\begin{array}{cc} {\bf 0} & \mathbcal{B}_{(w)} \\ \mathbcal{B}_{(w)}^{\dag} & {\bf 0}\end{array}\right),
\label{Dw3d4}
\eea
where $\mathbcal{B}_{(w)}$ is a matrix of dimension $2N\times 2M$ having a different definition depending on the type of the link. If we define the ``Pauli" matrices $\bm\sigma({\bf F})$ as
\bea
\hspace*{-8mm}\bm\sigma_1({\bf F})=\left(\begin{array}{cc}{\bf 0}&{\bf F} \\ {\bf F}&0\end{array}\right),\ \ \ \bm\sigma_2({\bf F})=\left(\begin{array}{cc}{\bf 0}&-\mathrm{i}{\bf F} \\ \mathrm{i}{\bf F} &{\bf 0}\end{array}\right),\ \ \bm\sigma_3({\bf F})=\left(\begin{array}{cc}{\bf F}& {\bf 0}\\{\bf 0}& -{\bf F}\end{array}\right),
\label{Pauli}
\eea
where ${\bf F}$ is a generic matrix of dimension $N\times M$,
we can express $\mathbcal{B}_{(w)}$ for $w\in \{x,y,z\}$ as 
\bea
\mathbcal{B}_{(x)}=\bm\sigma_1({\bf B}_{(x)}), \quad \mathbcal{B}_{(y)}=\bm{\sigma}_2({\bf B}_{(y)}),\quad\mathbcal{B}_{(z)}=\bm\sigma_3({\bf B}_{(z)}).
\eea
Having defined the directional Dirac operator $\bar\mathcal{D}$  in dimension $d=3$ (given by  Eq.(\ref{3dD})), we can formulate the  directional topological Dirac equation in $d=3$ as
\bea
i \partial_t\tilde{\bm\Psi}=\bar{\mathbcal{H}}\tilde{\bm\Psi}.
\label{Topo_Diracd3}
\eea
 with  Hamiltonian $\bar\mathbcal{H}$ given by 
\bea
\bar\mathbcal{H}=\bar\mathbcal{D}+m_0\bm\beta,
\label{H2}
\eea
where   $\bm\beta$ is now the matrix
\bea
\bm \beta=\left(\begin{array}{cc}{\bf I}_{2N} & {\bf 0}\\ {\bf 0}& -{\bf I}_{2M}\end{array}\right).
\label{beta3d}
\eea
We notice  that $\bm\beta$ anti-commutes with each $\mathbcal{D}_{(w)}$, i.e. 
\bea
\{\mathbcal{D}_{(w)},\bm\beta\}={\bf 0},
\eea
for $w\in \{x,y,z\}$
and the anti-commutators between the directional $d=3$-dimensional Dirac operators associated to different types of links $w\neq w'$ are given by  the  matrices
\bea
\{\mathbcal{D}_{(w)},\mathbcal{D}_{(w')}\}=\left(\begin{array}{cc} {\bf 0}& {\bf 0}\\ {\bf 0}&[\mathbcal{B}_{(w)}]^{\dag}\mathbcal{B}_{(w')}+{[\mathbcal{B}_{(w')}]^{\dag}}{\mathbcal{B}_{(w)}} \end{array}\right),
\eea
where $[\mathbcal{B}_{(w)}]^{\dag}\mathbcal{B}_{(w')}+{[\mathbcal{B}_{(w')}]^{\dag}}{\mathbcal{B}_{(w)}}$ can be expressed as
\bea
\mathbcal{B}_{(x)}^{\dag}\mathbcal{B}_{(y)}+\mathbcal{B}_{(y)}^{\dag}\mathbcal{B}_{(x)} =\mathrm{i}\epsilon_{123}\bm\sigma_{3}({\bf B}_{(x)}^{\dag}{\bf B}_{(y)}-{\bf B}_{(y)}^{\dag}{\bf B}_{(x)}),\nonumber\\
\mathbcal{B}_{(x)}^{\dag}\mathbcal{B}_{(z)}+\mathbcal{B}_{(z)}^{\dag}\mathbcal{B}_{(x)} =\mathrm{i}\epsilon_{132}\bm\sigma_{2}({\bf B}_{(x)}^{\dag}{\bf B}_{(z)}-{\bf B}_{(y)}^{\dag}{\bf B}_{(z)}),\nonumber\\
\mathbcal{B}_{(y)}^{\dag}\mathbcal{B}_{(z)}+\mathbcal{B}_{(z)}^{\dag}\mathbcal{B}_{(y)} =\mathrm{i}\epsilon_{231}\bm\sigma_{1}({\bf B}_{(y)}^{\dag}{\bf B}_{(z)}-{\bf B}_{(z)}^{\dag}{\bf B}_{(y)}),
\eea
with $\epsilon_{ijs}$ indicating the Levi-Civita symbol.
Similarly to what we have discussed for the  directional topological Dirac equation in dimension $d=2$, also the eigenstates of the  directional topological Dirac equation in dimension $d=3$ obey the dispersion relation 
\bea
E^2=m_0^2+\mu
\label{disd3}
\eea
where $\mu$ is the generic eigenvalue of the graph Laplacian ${\bf L}_{[0]}$ given by 
\bea
{\bf L}_{[0]}={\bf L}_{(x)}+{\bf L}_{(y)}+{\bf L}_{(z)}.
\eea
Here the $w$-type directional graph Laplacian  ${\bf L}_{(w)}$ is given by 
\bea
{\bf L}_{(w)}={\bf B}_{(w)}{\bf B}_{(w)}^{\dag},
\eea
with each pair of directional graph Laplacian commuting with each other, i.e.
\bea
[{\bf L}_{(w)},{\bf L}_{(w')}]={\bf 0}.
\eea
By indicating with $\lambda_w$ the generic eigenvalue of the incidence matrix ${\bf B}_{(w)}$ we have 
\bea
\mu=|\lambda_x|^2+|\lambda_y|^2+|\lambda_z|^2,
\eea
with
 \bea
 |\lambda_w|=2\sin(k_w/2),
 \eea 
 where  ${\bf k}=(k_x,k_y,k_z)$ indicates the wave-number and where  due to the periodic boundary conditions $k_w$ takes the values $k_w=2\pi \bar{n}_w/N^{1/3}$  with $\bar{n}_w$ integer  between $0$ and $N^{1/3}-1$.
  It follows that for $|{\bf k}|\ll1$, i.e. $\mu\ll1$ the density of eigenvalues $\rho(\mu)$ of the graph Laplacian follows Eq. (\ref{rhomuds}) and the density $g(E)$ of eigenstates of positive energy $E$ follows Eq. (\ref{gEds}) with $d_S=d=3$.
  In order to show that the eigenstates of the directional topological Dirac equation in $d=3$ follow the dispersion relation given by Eq. (\ref{disd3}), we indicate the  component ${\bf X}$ of the eigenstates as ${\bf X}=({\bf X}_{(x)},{\bf X}_{(y)},{\bf X}_{(z)})^{\top}$ where ${\bf X}_{(w)}$ is the component of ${\bf X}$ calculated on links of type $w$. With this choice of notation,  the directional topological Dirac  equations has
eigenstates associated to the  energy $E$  which satisfy
 \bea
 E\bm\Phi=\mathbcal{B}_{(x)}{\bf X}_{(x)}+\mathbcal{B}_{(y)}{\bf X}_{(y)}+\mathbcal{B}_{(z)}{\bf X}_{(z)}+m_0\bm\Phi,\nonumber\\
  E{\bf X}_{(x)}=[\mathbcal{B}_{(x)}]^{\dag}\bm\Phi-m_0{\bf X}_{(x)}, \nonumber\\
    E{\bf X}_{(y)}=[\mathbcal{B}_{(y)}]^{\dag}\bm\Phi-m_0{\bf X}_{(y)},\nonumber \\
      E{\bf X}_{(z)}=[\mathbcal{B}_{(z)}]^{\dag}\bm\Phi-m_0{\bf X}_{(z)}.
 \eea
 From this equation it follows that $\bm\Phi$  satisfies
 \bea
  (E-m_0)(E+m_0)\bm\Phi=\mathbcal{L}_{[0]}{\bm \Phi},
 \eea
 where $\mathbcal{L}_{[0]}$ is given by  
 \bea
 \mathbcal{L}_{[0]}=  \mathbcal{L}_{(x)}+  \mathbcal{L}_{(y)}+  \mathbcal{L}_{(z)},
 \eea
 and each directional $\mathbcal{L}_{(w)}$ can be written as
 \bea
  \mathbcal{L}_{(w)}=\mathbcal{B}_{(w)}[\mathbcal{B}_{(w)}]^{\dag}=\left(\begin{array}{cc}{\bf L}_{(w)} & {\bf 0}\\  {\bf 0}&{\bf L}_{(w)}\end{array}\right).
 \eea
 Therefore the  eigenstates of the directional topological Dirac equation in dimension $d=3$ have the   $\bm\Phi$ component proportional to the eigenvectors of the  Laplacian $\mathbcal{L}_{[0]}$. These eigenvectors (indicated here as $ \bm\Phi_i^{(r)}$ with $r\in \{1,2\}$)
  form a $d=3$-dimensional Fourier basis with associated wave-number ${\bf k}=(k_x,k_y,k_z)$ and for each value of the energy $E=\pm\sqrt{|\lambda_x|^2+|\lambda_y|^2+|\lambda_z|^2+m_0^2}$ have elements 
  \bea
  \Phi_i^{(1)}=\mathcal{N}\left(\begin{array}{c} 0\\1\end{array}\right) e^{\mathrm{i}{\bf k}\cdot {\bf r}_i},\quad  \Phi_i^{(2)}=\mathcal{N}\left(\begin{array}{c} 0\\1\end{array}\right) e^{\mathrm{i}{\bf k}\cdot {\bf r}_i},
  \eea
  where 
  ${\mathcal{N}}$ is a normalization constant.
 The component ${\bf X}^{(r)}=\left({\bf X}_{(x)}^{(r)},{\bf X}_{(y)}^{(r)},{\bf X}_{(z)}^{(r)}\right)^{\top}$ can be evaluated as  
  \bea
  {\bf X}_{(w)}^{(1)}=\frac{1}{E+m_0}\mathbcal{B}_{(w)}^{\dag}{\bm\Phi}^{(1)},\quad   {\bf X}_{(w)}^{(2)}=\frac{1}{E+m_0}\mathbcal{B}_{(w)}^{\dag}{\bm\Phi}^{(2)},
  \eea
  with $w\in \{x,y,z\}.$
Interestingly since ${\bf L}_{[0]}$ commutes with every directional ${\bf L}_{(w)}$, we obtain that ${\bf X}^{(r)}$ obtained in this way are  also eigenvectors of the $\mathbcal{L}_{[1],(w)}^{down}$ Laplacians given by 
\bea
\mathbcal{L}_{[1],(w)}^{down}=\mathbcal{B}_{(w)}^{\dag}\mathbcal{B}_{(w)},
\eea
for  $w\in \{x,y,z\}$.

\section{Topological  Dirac equation of multiplex  networks}
Interestingly the directional topological Dirac operator developed to treat the different types of links in  $d=3$ dimensional lattices, can be adapted to investigate the topological Dirac equation on multiplex networks\cite{bianconi2018multilayer}  as well. Here we consider the case of a duplex network, i.e. a multiplex network formed by two layers. Each layer of the network describes a different network of interactions, between the same set of nodes.
For instance in the brain network of C. elegans the two layer can represent electrical gap junctions or chemical synapses between the neurons.
In a single network each pair of nodes can be either connected or not connected by a link. In a duplex network instead each pair of nodes can be connected in multiple ways \cite{bianconi2013statistical}: they can be connected only in layer 1, they can be connected only in layer 2 or they can be connected in both layers. If they are only connected in layer 1, we say that they are connected by a multilink $\vec{m}=(1,0)$, if they are only connected in layer 2 we say that they are connected by a multilink $\vec{m}=(0,1)$, if they are connected in both layers we say they are connected by a multilink $(1,1)$ (see Figure $\ref{fig:5}$). 
Note that each pair of nodes can be connected only by a single type of multilink. 
This decomposition of multilinks of different types can be compared with the treatment of  different types of links  of the    $d=3$ dimensional lattice discussed  in the previous section.
This allows us to extend to duplex networks the formalism developed to treat the directional topological Dirac equation for $d=3$ dimensional lattices. 

Let us indicate with $N$ the total number of nodes of the duplex network and with $M$ the total number of multilinks of any type $\vec{m}\in \{(1,0),(0,1),(1,1)\}$. In order to treat the directional topological  Dirac equation on duplex networks, we decompose the  incidence matrix ${\bf B}$ of size $N\times M$ in three terms depending on the three types of multilinks
\bea
{\bf B}={\bf B}_{(1,0)}+{\bf B}_{(0,1)}+{\bf B}_{(1,1)}
\eea
 where here ${\bf B}_{\vec{m}}$ with $\vec{m}\in \{(1,0),(0,1),(1,1)\}$ describes the incidence matrix between nodes and multilinks of type $\vec{m}$, i.e. 
\bea
[{\bf B}_{\vec{m}}]_{i\ell}=\left\{\begin{array}{cccc}1 & \mbox{if}\   \ell=[j,i], &\mbox{and } \ell\  \mbox{is a  multilink of type $\vec{m}$ }\\ -1 & \mbox{if} \ \ell=[i,j],  &\mbox{and } \ell\  \mbox{is a  multilink of type $\vec{m}$ }\\  0 & \  \mbox{otherwise}. &\end{array}\right.
\label{Bdir}
\eea
This decomposition allows us to   consider the  directional Dirac operator $\bar{\mathbcal{D}}$
\bea
\bar\mathbcal{D}=\mathbcal{D}_{(1,0)}+\mathbcal{D}_{(0,1)}+\mathbcal{D}_{(1,1)}
\label{D2multi}
\eea
where the $\vec{m}$ directional Dirac operator is defined as
\bea
\mathbcal{D}_{\vec{m}}=\left(\begin{array}{cc} {\bf 0} & \mathbcal{B}_{\vec{m}} \\ \mathbcal{B}_{\vec{m}}^{\dag} & {\bf 0}\end{array}\right),
\eea
with $\mathbcal{B}_{\vec{m}}$ indicating the $2N\times 2M$ matrix given for $\vec{m}=(1,0)$, $\vec{m}=(0,1)$ and $\vec{m}=(1,1)$ by 
\bea
\mathbcal{B}_{(1,0)}=\bm\sigma_1({\bf B}_{(1,0)}), \quad \mathbcal{B}_{(0,1)}=\bm{\sigma}_2({\bf B}_{(0,1)}),\quad\mathbcal{B}_{(1,1)}=\bm\sigma_3({\bf B}_{(1,1)}),
\eea
with $\bm\sigma_s({\bf F})$ with $s\in \{1,2,3\}$  defined in Eq. (\ref{Pauli}).
Following the mapping to the directional Dirac equation in $d=3$ lattices we can define the  directional topological Dirac equation for duplex networks as in  Eq. (\ref{Topo_Diracd3}) with Hamiltonian given by Eq. (\ref{H2}) where $\bar{\mathbcal{D}}$ is given by Eq. (\ref{D2multi}).
Like for the directional topological Dirac operator in $d=3$ dimensions, the eigenstates of the directional topological Dirac equation for duplex networks admits eigenstates that obey the dispersion relation 
\bea
E^2=m_0^2+\mu,
\eea
where $\mu$ is the eigenvalue of the graph Laplacian ${\bf L}_{[0]}$ given by 
\bea
{\bf L}_{[0]}={\bf L}_{(0,1)}+{\bf L}_{(1,0)}+{\bf L}_{(1,1)},
\eea
with ${\bf L}_{\vec{m}}$ given by 
\bea
{\bf L}_{\vec{m}}={\bf B}_{\vec{m}}{\bf B}_{\vec{m}}^\dag.
\eea
However for duplex networks,  we have that   a very notable difference with respect to the $d=3$ dimensional lattices. Indeed the Laplacians corresponding to distinct multilinks in the vast majority of cases do not commute, i.e.
\bea
[{\bf L}_{\vec{m}},{\bf L}_{\vec{m}'}]\neq {\bf 0},
\label{Lnc}
\eea
if $\vec{m}'\neq\vec{m}$.
Therefore the eigenstates $(\bm\Phi^{(r)},{\bf X}^{(r)})$ with $r\in \{1,2\}$, corresponding to the energy $E=\pm\sqrt{\mu+m_0^2}$ have the $\bm\Phi^{(r)}$ component with elements 
 \bea
  \Phi_i^{(1)}=\mathcal{N}\left(\begin{array}{c} 0\\1\end{array}\right) { u}^{\mu}_i,\quad  \Phi_i^{(2)}=\mathcal{N}\left(\begin{array}{c} 0\\1\end{array}\right) { u}^{\mu}_i,
  \eea
  where ${\bf u}^{\mu}$ is the eigenvector of the graph Laplacian ${\bf L}_{[0]}$ corresponding to the eigenvalue $\mu$ and  where ${\mathcal{N}}$ is a normalization constant.
Additionally the ${\bf X}^{(r)}$ component  of the mentioned eigenstates admit the decomposition 
${\bf X}^{(r)}=\left({\bf X}_{(1,0)}^{(r)},{\bf X}_{(1,0)}^{(r)},{\bf X}_{(1,1)}^{(r)}\right)^{\top}$ with
\bea
{\bf X}_{\vec{m}}^{(r)}=\frac{1}{E+m_0}\mathbcal{B}_{\vec{m}}^{\dag}{\bm\Phi}^{(r)},
\eea
However  since the graph Laplacians associated to different types of multilinks do not commute,  i.e. Eq. (\ref{Lnc}) applies, the component ${\bf X}$ of the eigenstates of the directional topological Dirac equation on duplex networks is not an eigenvector of the Laplacians
\bea
\mathbcal{L}_{[1],\vec{m}}^{down}=\mathbcal{B}_{\vec{m}}^{\dag}\mathbcal{B}_{\vec{m}},
\eea
for  any type of multilink $\vec{m}\in \{(1,0),(0,1),(1,1)\}$.

In Table~\ref{table:1}, Table~\ref{table:2} and Table~\ref{table:3} we evaluate to what extent the multi-Laplacians ${\bf L}_{\vec{m}}$ associated to different multilinks of real multiplex networks do not commute. The metric used to evaluate the deviation from zero of the commutator $[{\bf L}_{\vec{m}},{\bf L}_{\vec{m}'}]$ is  the Euclidean metric given by the square-root of the normalized sum of the squares of the elements of the matrix.
The data that we consider includes the collaboration network of scientists working in Network Science, the C.elegans duplex connectome, the duplex  network of the Arabidopsis genetic layers. The Euclidean metric measured on the commutators $[{\bf L}_{\vec{m}},{\bf L}_{\vec{m}'}]$ display a wide range of values with the C.elegans duplex network displaying the largest values indicating the most significant deviation from the commutating scenario.

\begin{table}
\caption{\label{table:1}
The Euclidean metric of the commutator between the Laplacian ${\bf L}_{\vec{m}}$ and the Laplacian ${\bf L}_{\vec{m}'}$ for a duplex networks of collaboration between scientists working in  Network Science. The duplex network is constructed  from the dataset published in  \cite{de2015identifying} with the first layer including collaborations or works published in the arxiv reporsitory 
physics.soc-ph while the second layer includes collaborations of work published in the arxiv repository cond-mat.dis-nn.
The total number of nodes is $N=4537$, the total number of pairs of nodes connected in both layers is $O=650$. }
\begin{tabular*}{\textwidth}{@{}l*{15}{@{\extracolsep{0pt plus
12pt}}l}}
\br
$\vec{m}$/$\vec{m}'$&$(1,0)$&$(0,1)$&$(1,1)$\\
\mr
$(1,0)$&$0$&$0.112$ &$0.030$\\
$(0,1)$&$0.112$&$0$&$0.1057$\\
$(1,1)$&$0.030$&$0.1057$&$0$\\
\br
\end{tabular*}
\end{table}

\begin{table}
\caption{\label{table:2}
The Euclidean metric of the commutator between the Laplacian ${\bf L}_{\vec{m}}$ and the Laplacian ${\bf L}_{\vec{m}'}$ for a duplex networks of gap-junction and synaptic connections in C. elegans.. The duplex network is constructed  from the dataset published in  \cite{chen2006wiring} and curated in  \cite{de2015muxviz} with the first layer including gap-junctions and the second layer including monosynaptics connections. The networks in each layers are symmetrized and their tadpoles have been disregarded. 
The total number of nodes is $N=275$, the total number of pairs of nodes connected in both layers is $O=111$. }
\begin{tabular*}{\textwidth}{@{}l*{15}{@{\extracolsep{0pt plus
12pt}}l}}
\br
$\vec{m}$/$\vec{m}'$&$(1,0)$&$(0,1)$&$(1,1)$\\
\mr
$(1,0)$&$0$&$1.453$ &$0.825$\\
$(0,1)$&$1.453$&$0$&$1.051$\\
$(1,1)$&$0.825$&$1.051$&$0$\\
\br
\end{tabular*}
\end{table}

\begin{table}
\caption{\label{table:3}
The Euclidean metric of the commutator between the Laplacian ${\bf L}_{\vec{m}}$ and the Laplacian ${\bf L}_{\vec{m}'}$ for a duplex networks capturing Arabidopsis genetic layers. The duplex network is constructed  from the dataset published in  \cite{stark2006biogrid} and curated in  \cite{de2015structural} with the first layer including direct interactions the second layer physical association. 
The total number of nodes is $N=6903$, the total number of pairs of nodes connected in both layers is $O=628$.  }
\begin{tabular*}{\textwidth}{@{}l*{15}{@{\extracolsep{0pt plus
12pt}}l}}
\br
$\vec{m}$/$\vec{m}'$&$(1,0)$&$(0,1)$&$(1,1)$\\
\mr
$(1,0)$&$0$&$0.620$ &$0.120$\\
$(0,1)$&$0.620$&$0$&$0.076$\\
$(1,1)$&$0.120$&$0.076$&$0$\\
\br
\end{tabular*}
\end{table}

\section{Directional Topological  Dirac equation in  $1+d$-dimensional space-time lattice}
\subsection{Directional Topological  Dirac equation in  $1+d$-dimensional space-time lattice with $d\in \{2,3\}$}

A very important question is whether we can generalize the directional  topological Dirac equation to treat differently  space-like and time-like links in a $1+d$-dimensional space-time.
This operation  entails considering a directional topological Dirac equation in which also time is discretized.
To start investigating this question we consider a $1+d$ dimensional  lattice with $d\in\{1,2\}$ representing the entire space-time with $d$ dimensions indicating space and an additional  dimension indicating time, so that the links will be distinguished between $x$-type, $y$-type links and $t$-type links.
For simplicity we assume the space is isotropic portion of the lattice in $\mathbb{R}^{d+1}$  with sides of length ${N}^{1/(d+1)}$ and we consider period boundary conditions.
For $1+1$ lattice incidence matrix ${\bf B}$ can be decomposed as
\bea
{\bf B}={\bf B}_{(x)}+{\bf B}_{(t)}
\eea
for $1+2$ lattice the incidence matrix can be decomposed as 
\bea
{\bf B}={\bf B}_{(x)}+{\bf B}_{(y)}+{\bf B}_{(t)}
\eea
where ${\bf B}_{(w)}$ with $w\in \{x,y,t\}$ indicates the directional incidence matrix   defined by Eq. (\ref{Bdir}) .

Starting from these directional incidence matrices we can define the directional Dirac operator $\bar\mathbcal{D}$ as 
\bea
\bar{\mathbcal{D}}=\sum_{w}\mathbcal{D}_{(w)}
\eea
where for $1+1$ lattices the sum extends to $w\in \{x,y\}$ and for $1+2$ lattices it extends to $w\in \{x,y,t\}$. Here   the $w$-directional Dirac operators $\mathbcal{D}_{(w)}$ corresponding to the spatial directions $w\in \{x,y\}$ have the same defintion as for $d$ dimensional spatial lattices, and are given by  Eq. (\ref{Ddir2d}) with the directional Dirac operator $\mathbcal{D}_{(t)}$ corresponding to the temporal direction is  defined as  
\bea
 \mathbcal{D}_{(t)}=\left(\begin{array}{cc}0& \mathrm{i}{\bf B}_{(t)}\\\mathrm{i}{[{\bf B}_{(t)}]}^{\dag}&0\end{array}\right).
\eea
We notice that while $\mathbcal{D}_{(x)}$ and $\mathbcal{D}_{(y)}$  are Hermitian, we have chosen $\mathbcal{D}_{(t)}$ to be anti-Hermitian. With this choice of the  $t$-directional Dirac operator $\mathbcal{D}_{(t)}$, adopting for  $\bm\beta$ the definition  given by Eq. (\ref{beta0}), one observes that the anti-commutator between $\bm\beta$ and $\mathbcal{D}_{(w)}$ vanishes, i.e.
\bea
\{\mathbcal{D}_{(w)},\bm\beta\}={\bf 0},
\eea
for $w\in \{x,y,t\}$.
However the anti-commutator between $\mathbcal{D}_{(w)}$ with $w\in \{x,y\}$ and $\mathbcal{D}_{(t)}$ is given by 
 \bea
 \{\mathbcal{D}_{(w)},\mathbcal{D}_{(t)}\}=\left(\begin{array}{cc} {\bf 0}& {\bf 0}\\ {\bf 0}&\mathrm{i}\left([{\bf B}_{(w)}]^{\dag}{\bf B}_{(t)}+{[{\bf B}_{(t)}]^{\dag}}{{\bf B}_{(w)}} \right)\end{array}\right).
 \eea
 The directional topological Dirac equation on this $1+d$ space-time is given by the eigenvalue problem
 \bea
 \left(\bar{\mathbcal{D}}+m_0\bm\beta\right)\bm\psi=0
 \eea
 with the topological spinor  ${\bm \psi}=(\bm\phi,\bm\chi)^{\top}$ defined in Eq.(\ref{spinor}).
 Let us study this equation in the case of a $1+1$ dimensional lattice (as the case $1+2$ is a straightforward generalization).
 By writing this eigenvalue problem for the component $\bm\phi$ and the component $\bm\chi$ separately, where the component $\bm\chi$ is decomposed according to the type of link in which the $1$-cochain is defined, $\bm\chi=(\bm\chi_{(x)},\bm\chi_{(t)})^{\top}$  we obtain
 \bea
 {\bf B}_{(x)}\bm\chi_{(x)}+\mathrm{i}{\bf B}_{(t)}\bm\chi_{(t)}+m_0\bm\phi={\bf 0},\nonumber \\
  {\bf B}_{(x)}^{\dag}\bm\phi-m_0\bm\chi_{(x)}={\bf 0},\nonumber \\
   \mathrm{i}{\bf B}_{(t)}^{\dag}\bm\phi-m_0\bm\chi_{(t)}={\bf 0}.
 \eea
 From this system of  equations it is possible to observe that $\bm\phi$ must be an eigenvector of the discrete D'Alembert operator 
 \bea
 {\mathbf{\square}}={\bf L}_{(t)}-{\bf L}_{(x)}={\bf B}_{(t)}{\bf B}_{(t)}^{\dag}-{\bf B}_{(x)}{\bf B}_{(x)}^{\dag}
 \eea
 with eigenvalue $m_0^2$.
 Therefore  the component $\bm\phi$ of the topological spinor calculated on a node $i$ of coordinates $(t_i,x_i)$ is given by 
 \bea
 \phi_i=\mathcal{N}e^{-\mathrm{i}( \omega t_i-k_x x_i)}
 \eea
 where $\mathcal{N}$ is a normalization constant. Here $\omega=2\pi\bar{n}_\omega/N^{1/2}, k_x=2\pi\bar{n}_x/\sqrt{N}$ with $0\leq \bar{n}_w\leq N^{1/2}-1$. Let us   indicate with $\lambda_x$ and $E_t$ the eigenvalues of the directional incidence matrices ${\bf B}_{(x)}$ and ${\bf B}_{(t)}$  with 
 with 
 \bea
 |\lambda_w|=2\sin(k_w/2),
 \eea
 for $w\in \{x,y\}$.
 In order to satisfy the directional topological Dirac equation on $1+1$ space-time $E_t$ and $\lambda_x$ must satisfy the dispersion relation 
 \bea
 |E_t|^2=|\lambda_x|^2+m_0^2,
 \eea 
 which, given the discrete nature of the spectrum can impose constraints on the possible value of the mass $m_0$ and eigenvalues of ${\bf B}_t$ and ${\bf B}_x$ that might be considered. 
 If this relation can be satisfied the eigenstate exists, and the topological Dirac equation on discrete space time has a solution with  the components  $\bm\chi_{(x)}$ and $\bm\chi_{(y)}$  given by 
 \bea
 \bm\chi_{(x)}=\frac{1}{m_0}{\bf B}_{(x)}^{\dag}\bm\phi,\nonumber \\
 \bm\chi_{(t)}=\mathrm{i}\frac{1}{m_0}{\bf B}_{(t)}^{\dag}\bm\phi.
 \eea 
By following similar steps it is possible to show that for the $1+2$ topological Dirac equation, the component $\bm\phi$ is an eigenvector of the operator 
\bea
 {\mathbf{\square}}={\bf L}_{(t)}-{\bf L}_{(x)}-{\bf L}_{(y)}={\bf B}_{(t)}{\bf B}_{(t)}^{\dag}-{\bf B}_{(x)}{\bf B}_{(x)}^{\dag}-{\bf B}_{(y)}{\bf B}_{(y)}^{\dag}
\eea
 with eigenvalue $m_0^2$ and its element $\phi_i$ calculated  
 on a node $i$ of coordinates $(t_i,x_i,y_i)$ is given by 
 \bea
 \phi_i=\mathcal{N}e^{-\mathrm{i}(\omega t_i-k_x x_i-k_y y_i)}
 \eea
 where $\mathcal{N}$ is a normalization constant and $k_w=2\pi\bar{n}_w/N^{1/3}$ with $0\leq \bar{n}_w\leq N^{1/3}-1$. Indicating with  $\lambda_x$, $\lambda_y$ and $E_t$ the eigenvalues of the directional incidence matrices ${\bf B}_{(x)}$, ${\bf B}_{(y)}$ and ${\bf B}_{(t)}$ respectively we see that for $1+2$ dimensional space-time this eigenvalues must satisfy the dispersion relation 
 \bea
 |E_t|^2=|\lambda_x|^2+|\lambda_y|^2+m_0^2.
 \eea
 Therefore, due to the discrete nature of the spectrum, solving for the spectrum of the directional topological Dirac equation  reduces to a problem connected to number theory. 
 
\subsection{Directional Topological Dirac equation on $1+3$ dimensional space-time lattice}
In this section we further extend the directional topological Dirac equation to $1+3$ dimensions.
To this end we consider the topological spinor $\bm\Psi=(\bm\Phi,{\bf X})^{\top}$ defined in Eq. (\ref{spinord3}).

The incidence matrix ${\bf B}$ can be written as a sum of directional incidence matrix ${\bf B}_{(w)}$ with $w\in\{x,y,z,t\}$.
The directional Dirac operators $\mathbcal{D}_{(w)}$ with $w\in \{x,y,z,\}$ have the same definition as for $d=3$ spatial lattices given by Eq. (\ref{Dw3d4}). The directional Dirac operator in the time direction is defined instead as 
\bea
\mathbcal{D}_{(t)}=\left(\begin{array}{cc} {\bf 0} & \mathbcal{B}_{(t)} \\ -\mathbcal{B}_{(t)}^{\dag} & {\bf 0}\end{array}\right),
\label{Dt4d}
\eea
with $\mathbcal{B}_{(t)}$ given by 
\bea
\mathbcal{B}_{(t)}=\left(\begin{array}{cc} \mathrm{i}{\bf B}_{(t)}&{\bf 0}  \\  {\bf 0}& \mathrm{i}{\bf B}_{(t)}\end{array}\right).
\eea
We notice that if we define $\bm\beta$ according to Eq. (\ref{beta3d}), we have 
\bea
\{\mathbcal{D}_{(t)},\bm\beta\}={\bf 0}.
\eea
Moreover the anti-commutators between the spatial Dirac operators and the temporal Dirac operator are given by  
\bea
\{\mathbcal{D}_{(w)},\mathbcal{D}_{(t)}\}=\left(\begin{array}{cc} {\bf 0}& {\bf 0}\\ {\bf 0}&[\mathbcal{B}_{(w)}]^{\dag}\mathbcal{B}_{(t)}-{[\mathbcal{B}_{(t)}]^{\dag}}{\mathbcal{B}_{(w)}} \end{array}\right),
\eea
where $[\mathbcal{B}_{(w)}]^{\dag}\mathbcal{B}_{(t)}-{[\mathbcal{B}_{(w')}]^{\dag}}{\mathbcal{B}_{(t)}}$ can be expressed as
\bea
\mathbcal{B}_{(x)}^{\dag}\mathbcal{B}_{(t)}-\mathbcal{B}_{(t)}^{\dag}\mathbcal{B}_{(x)} =\mathrm{i}\bm\sigma_{1}({\bf B}_{(x)}^{\dag}{\bf B}_{(t)}+{\bf B}_{(t)}^{\dag}{\bf B}_{(x)}),\nonumber\\
\mathbcal{B}_{(y)}^{\dag}\mathbcal{B}_{(t)}-\mathbcal{B}_{(t)}^{\dag}\mathbcal{B}_{(y)} =\mathrm{i}\bm\sigma_2({\bf B}_{(y)}^{\dag}{\bf B}_{(t)}+{\bf B}_{(t)}^{\dag}{\bf B}_{(y)}),\nonumber\\
\mathbcal{B}_{(z)}^{\dag}\mathbcal{B}_{(t)}-\mathbcal{B}_{(z)}^{\dag}\mathbcal{B}_{(y)} =\mathrm{i}\bm\sigma_{3}({\bf B}_{(y)}^{\dag}{\bf B}_{(z)}+{\bf B}_{(z)}^{\dag}{\bf B}_{(y)}),
\eea
where $\bm\sigma({\bf F})$ are given by Eq. (\ref{Pauli}).
The directional topological Dirac equation on a $1+3$ dimensional space-time can therefore we written as 
\bea
(\bar{\mathbcal{D}}+m_0\bm\beta)\bm\Psi={\bf 0},
\eea
where the directional Dirac operator $\bar{\mathbcal{D}}$ is given by 
\bea
\bar{\mathbcal{D}}=\sum_{w\in \{x,y,z,t\}}\mathbcal{D}_{(w)}.
\eea
The study of this equation shows that the topological spinor $\bm\Psi=(\bm\Phi,{\bf X})^{\top}$ has the component $\bm\Phi$ which is an eigenvector of the discrete D'Alembert operator 
\bea
 {\mathbf{\square}}=\mathbcal{L}_{(t)}-\mathbcal{L}_{(x)}-\mathbcal{L}_{(y)}-\mathbcal{L}_{(z)}
\eea
 with eigenvalue $m_0^2$, where $\mathbcal{L}_{(w)}$ are given by 
 \bea
 \mathbcal{L}_{(z)}=\left(\begin{array}{cc} {\bf B}_{(w)}{\bf B}_{(w)}^{\dag}&{\bf 0}  \\  {\bf 0}& {\bf B}_{(w)}{\bf B}_{(w)}^{\dag}\end{array}\right),
 \eea
 for $w\in\{t,x,y,z\}$.
 The eigenvectors of $\square$ are eigenvector of the spatial $\mathcal{B}_{(w)}$ with eigenvalues $\lambda_w$ and of the temporal $\mathbcal{B}_{(t)}$ with eigenvalue $E_t$ where $E_t$ and $\lambda_w$ which must satisfy
 \bea
 E_t^2=|\bm\lambda|^2+m_0
 \eea
 where $\bm\lambda=(\lambda_x,\lambda_y,\lambda_z)$.
 Given that we have assumed to have a finite $1+3$ space-time with periodic boundary conditions, $\lambda_w$ and $E_t$ can only take discrete values fixed by the pair  $(\omega,{\bf k})$ with 
 \bea
 |\lambda_w|=2\sin (k_w/2)|\quad |E_t|=2\sin(\omega/2)
 \eea
 with $k_w$ and $\omega$ quantized and given by $k_w=2\pi \bar{n}_w/N^{1/4}$, $\omega=2\pi\bar{n}_{\omega}/N^{1/4}$ with $\bar{n}_w$ and $\bar{n}_{\omega}$ taking integer values between zero and $N^{1/4}-1$.
 The component $\bm\Phi^{(r)}$ with $r\in \{1,2\}$ of the eigenstates of energy $E_t$ has elements 
 \bea
  \Phi_i^{(1)}=\mathcal{N}\left(\begin{array}{c} 0\\1\end{array}\right) e^{-\mathrm{i}(\omega t_i-{\bf k}\cdot {\bf r}_i)},\quad  \Phi_i^{(2)}=\mathcal{N}\left(\begin{array}{c} 0\\1\end{array}\right) e^{-\mathrm{i}(\omega t_i-{\bf k}\cdot {\bf r}_i)},
 \eea
where we have indicated with ${\bf k}=(k_x,k_y,k_z)$ and components ${\bf X}^{(r)}=\left({\bf X}_{(x)}^{(r)},{\bf X}_{(y)}^{(r)},{\bf X}_{(z)}^{(r)},{\bf X}_{(t)}^{(r)}\right)^{\top}$ given by 
 \bea
   {\bf X}_{(w)}^{(1)}=\frac{1}{m_0}\mathbcal{B}_{(w)}^{\dag}{\bm\Phi}^{(1)},\quad   {\bf X}_{(w)}^{(2)}=\frac{1}{m_0}\mathbcal{B}_{(w)}^{\dag}{\bm\Phi}^{(2)},
 \eea
 for the spatial directions $w\in \{x,y,z\}$ and  for the temporal direction $t$, ${\bf X}_{(t)}^{(r)}$ are:
 \bea
  {\bf X}_{(t)}^{(1)}=\mathrm{i}\frac{1}{m_0}\mathbcal{B}_{(t)}^{\dag}{\bm\Phi}^{(1)},\quad   {\bf X}_{(t)}^{(2)}=\mathrm{i}\frac{1}{m_0}\mathbcal{B}_{(t)}^{\dag}{\bm\Phi}^{(2)},
 \eea 

\section{Conclusions}
In this work we have formulated the topological Dirac equation and its directional variations and we have studied the properties of this wave equation on networks (simple and multilayer) and on simplicial complexes.
The main difference between the Dirac equation and the topological Dirac equation is that the wave function of the topological Dirac equation is discretized and defined on nodes, links and higher dimensional simplices. Therefore the topological spinor has a strict topological interpretation and relaxes the requirement of the locality of the wave function to the extent that it includes for instance 1-cochains defined on the links of the network.
We have shown that the topological Dirac equation obeys a relativistic dispersion relation in which the eigenvalue of the Dirac operator plays the role of the momentum.
The topological Dirac operator has been  extended to treat also simplicial complexes. In this context we observe that the eigenfunctions of the higher-order topological Dirac operator can be localized only on $n$ and $n-1$ dimensional simplices and that  correspondingly the dispersion relation is described by independent energy bands associated to the spectrum of the $n$-order up Laplacian.  

While the topological Dirac equation is defined on an arbitrary topology, we have considered a variation of this equation, the directional Dirac equation on $d=2$ and $d=3$ dimensional lattices and multiplex networks. The directional equation treats interactions of different types differently introducing a rotation of the spinor that is a function of the type of the interaction. While for lattices the type of interaction distinguishes between the different orthogonal directions of the links on multiplex networks the interactions of different types indicate different types of multilinks.
Therefore we see that if we want to distinguish between different multilinks of multiplex networks the directional topological Dirac equation naturally leads to  rotations of the topological spinor.

Finally we have discussed the  extension of this work to treat a discrete network  describing the entire space-time topology over which the directional topological Dirac equation can be defined.
In particular we formulate the directional topological Dirac equation defined on $1+d$ lattices with $d\in \{1,2,3\}$ spatial dimensions. This sheds some light on the notoriously difficult problem of defining a Lorentzian metric for a discrete geometry. Interestingly here the lattice is treated as an Euclidean lattice in $\mathbb{R}^{1+d}$ and the Loretzian nature of the lattice is only due to the opportune definition of the directional spatial and temporal Dirac operators.

 This work can be expanded in different directions.  From a Network Science perspective, it is interesting to study also the dissipative dynamics described by the Dirac operator, this would entail making a Wick rotation of the time in the topological Dirac equation. Research in this direction would  be  important to investigate whether the Dirac operator could be used for signal processing of different topological signals simultaneously and whether it could be explored as an important operator to define convolution of simplicial neural networks. 
 
 Many other research questions remain open if one desires to push forward the mapping between this mathematical framework  and physics systems. One important question that arises is whether one can observe symmetry breaking and the emergence of a renormalized mass from the study of non-linear topological Dirac equations similarly to what happens for  the non-linear Dirac equation captured by the  Nambu Jona-Lasinio model \cite{nambu1961dynamical}.  In this case the renormalized mass might acquire a different value depending on the spectral properties of the network or of the simplicial complex expanding our understanding of the relation between discrete topology, geometry and dynamics.
Another important direction that could be explored is the introduction of appropriate topological fields that might induce interaction terms able to describe transitions between different eigenstates  of the  topological Dirac operator. These terms could eventually lead to transition across different bands of the topological Dirac operator defined on simplicial complexes.
Finally the directional topological Dirac equation defined on discrete-space time could be extended to general discrete networks and  simplicial complex geometries.

We hope that these results will inspire further applied and theoretical work along these lines  revealing the rich interplay between the topology of networks and simplicial complexes and the evolution of wave functions and topological signals defined on them.

 
\section*{Data availability.}
The fungi data  published in \cite{Lee_2016_CP} is freely available at the repository \cite{fungi}.
The multiplex network datasets used in this work are all freely available at the repository \cite{manlio}.

\section*{Code availability.}
The code to generate the simplicial complex ``Network Geometry with Flavor" \cite{bianconi2016network,bianconi2017emergent} is freely available at the repository \cite{gin_git}. All other codes used in this work are available upon request.
\section*{References}
\bibliographystyle{unsrt}
\bibliography{references}
\end{document}